# Dynamical Configurations of Celestial Systems Comprised of Multiple Irregular Bodies

Yu Jiang[1,2], Yun Zhang[2], Hexi Baoyin[2], Junfeng Li[2]

1. State Key Laboratory of Astronautic Dynamics, Xi'an Satellite Control Center, Xi'an 710043, China

2. School of Aerospace Engineering, Tsinghua University, Beijing 100084, China

Y. Jiang (✉) e-mail: jiangyu_xian_china@163.com (corresponding author)

**Abstract**. This manuscript considers the main features of the nonlinear dynamics of multiple irregular celestial body systems. The gravitational potential, static electric potential, and magnetic potential are considered. Based on the three established potentials, we show that three conservative values exist for this system, including a Jacobi integral. The equilibrium conditions for the system are derived and their stability analyzed. The equilibrium conditions of a celestial system comprised of $n$ irregular bodies are reduced to $12n - 9$ equations. The dynamical results are applied to simulate the motion of multiple-asteroid systems. The simulation is useful for the study of the stability of multiple irregular celestial body systems and for the design of spacecraft orbits to triple-asteroid systems discovered in the solar system. The dynamical configurations of the five triple-asteroid systems 45 Eugenia, 87 Sylvia, 93 Minerva, 216 Kleopatra, and 136617 1994CC, and the six-body system 134340 Pluto are calculated and analyzed.

## 1. Introduction

Nonlinear dynamical behaviors in space science have been analyzed theoretically and numerically in previous studies, which include behaviors such as periodic orbits in three-dimensional (3D) potentials (Zotos and Caranicolas 2012), escape dynamics in galaxies (Zotos 2015), equilibria in binary-asteroid systems (Vera 2009a; Jacobson and Scheeres 2011) or the three body problem (Vera 2009b), stability of planetary rings (Salo and Schmidt 2010), and the instability regions of moonlets for the triple-body system 87 Sylvia (Frouard et al. 2012). However, most of the multiple asteroid systems in the Solar system are comprised of irregular minor bodies (Lagerros 1997; Beauvalet and Marchis 2014). The nonlinear dynamical study of multiple asteroid systems should not only consider the orbit of each body, but also consider the shape and attitude of each body (Marchis et al. 2012). For instance, Brozović et al. (2011) analyzed the motion of triple near-Earth Asteroid 1994 CC, and suggested that the attitude motion and orbit motion of the inner satellite are synchronous, which would indicate that the satellite and the primary are spin-orbit locked. In this paper we study the nonlinear dynamical behaviors of celestial systems comprised of multiple irregular bodies.

Since the discovery of the first triple-asteroid system 87 Sylvia in 2005 (Marchis et al. 2005; Frouard and Compère 2012; Berthier et al. 2014), eight additional triple-asteroid systems have been discovered in the solar system including 45 Eugenia (Marchis et al. 2010; Beauvalet et al. 2011; Beauvalet et al. 2012), 93 Minerva (Marchis et al. 2011, 2013), 216 Kleopatra (Descamps et al. 2011), 3749 Balam (Vokrouhlický et al. 2009; Marchis et al. 2012), 47171 1999TC$_{36}$ (Stansberry et al. 2006; Benecchi et al. 2010), 136108 Haumea (Pinilla-Alonso et al. 2009; Dumas et al. 2011), 136617 1994CC (Brozović et al. 2010, 2011; Fang et al. 2011), and 153591 2001SN$_{263}$ (Fang et al. 2011; Araujo et al. 2012; Sarli et al. 2012). With the exception of 3749 Balam and 136617 1994CC, the remaining seven are high size ratio triple-asteroid systems (Marchis et al. 2005, 2010, 2011, 2012; Pinilla-Alonso et al. 2009; Benecchi et al. 2010; Descamps et al. 2011; Fang et al. 2011). The irregular shapes and gravitational fields of binary-asteroid systems have been modeled by the two-polyhedron method (Werner and Scheeres 2005; Hirabayashi and Scheeres 2013). In addition, the discrete element method has been applied to binary systems (Mindlin and Deresiewica 1953; Ghaboussi and Barbosa 1990; Mishra and Rajamani 1992; Richardson et al. 1998; Walsh and Richardson 2006; Richardson et al. 2009), and is also appropriate for modeling celestial systems comprised of a greater number of irregular bodies.

The dynamics of multiple irregular-body systems exhibit both orbital motion and attitude motion, and each body in the system is acted upon by resultant gravitational forces and torques. Maciejewski (1995) investigated the dynamical model and relative equilibria of the gravitational two-body problem, and presented the gravitational potential energy, the gravitational force, and the resultant gravitational torque of the two-body problem. Koon et al. (2004) discussed the dynamical equation of an asteroid pair, and employed geometric mechanics to obtain a description of the system's reduced phase space. Gabern et al. (2006) studied spacecraft motion near a binary asteroid and found quasi-periodic orbits for parking the spacecraft while it is observing the asteroid pair. The dynamical equations of an asteroid pair can also be derived via Lagrangian and Hamiltonian reductions and expressed in the discrete form as a Lie group variational integrator that is suitable for numerical simulations (Cendra and Marsden 2005; Lee et al. 2007).

The stability condition for the two gravitationally interacting celestial bodies of the full two-body problem was presented in Scheeres (2002). The Hill unstable cases can be preliminarily modeled using a triaxial ellipsoid model (Scheeres 2004). The motion stability of the two moonlets of 87 Sylvia were discussed with respect to the second order coefficient of the nonspherical gravitational potential of the system's primary body (i.e., $J_2$), indicating that the moonlets are in a secular resonance and the longitudes of their nodes are locked (Winter et al. 2009). The internal region and the long-term stability in the internal region of the triple-asteroid system 2001 SN263 have been discussed (Araujo et al. 2012), indicating that the stable region is found to be in the neighborhoods of the moonlets.

The relative equilibria of high size ratio binary-asteroid systems can be modeled by an irregular body and a minor sphere (Scheeres 2006). The relative equilibria may be Lagrangian equilibria or non-Lagrangian equilibria (Maciejewski 1995). The necessary and sufficient conditions for stability of Lagrangian equilibria (Vera 2009a) and Eulerian equilibria (Vera 2009b) of binary asteroids can be given using geometric-mechanics methods. Vera and Vigueras (2006) studied the dynamics of a gyrostat in a gravitational system comprised of *n* spherical rigid bodies, and established some Lagrangian and Eulerian equilibria of the gyrostat in the four-body problem. Scheeres (2012) discussed the minimum energy configurations of the gravitational *n*-body problem. The primary of 216 Kleopatra has seven relative equilibria, which are equilibria for a massless particle in the potential of the primary (Jiang 2015; Jiang and Baoyin 2014; Jiang et al. 2014; Jiang et al. 2015a, 2015b, 2015c). If the parameters of the primary vary, the number, positions, and topological classifications of the relative equilibria may change (Jiang et al. 2015a, 2015b,

2015c).

The present study investigates the dynamics of multiple irregular celestial body systems by considering the gravitational potential, static electric potential, and magnetic potential. We derived formulas for the gravitational potential energy, static electric potential energy, and magnetic potential energy of a generalized $n$-body system. The total gravitational force, total static electric force, and total magnetic force acting on each body of the system are also derived. The resultant torque acting on each body has been presented. The dynamics equation relative to the inertial space and body-fixed frame of the primary asteroid of the group is presented. We establish that three conservative values exist for this system.

The relative equilibrium of the system is analyzed, the equilibrium conditions are derived, and the stability of the relative equilibrium is discussed. It is demonstrated that the equilibrium conditions of a celestial system comprised of $n$ bodies can be reduced to $12n - 9$ equations.

The results of the investigation are applied to the calculation of the orbital and attitude motion of the triple-asteroid systems 45 Eugenia, 87 Sylvia, 93 Minerva, 216 Kleopatra, and 136617 1994CC, as well as the six-body system 134340 Pluto, which includes the moonlet Charon. The amplitudes of the semi-major axes of the moonlets are found to be large for each of these triple-asteroid systems. The results show that all the moonlets of these triple-asteroid systems and those of the six-body system have small inclinations except for 93 Minerva. The moonlets of 93 Minerva, Aegis and Gorgoneion, have inclination values of 43.77 ° and 50.35 °, respectively. Among the 134340 Pluto system, although Pluto and Charon are gravitationally locked and continuously present the same face to each other, the orbital angular speeds and attitude angular speeds of Pluto and Charon are not constant, and exhibit a periodic variation. The figure of the relation diagram of $a$ and $e$ for moonlets of (93) Minerva and (136617) 1994CC are quite different from that for (45) Eugenia, (87) Sylvia, and (216) Kleopatra.

## 2. Dynamical Equation and Effective Potential

We consider a celestial system comprised of $n$ irregular-shaped bodies (see Figs. 1 and 2). We denote $\beta_l (l = i, j)$ as the $l$-th body, $G$ as the Newtonian gravitational constant, $\mathbf{r}_l$ as the position vector of the mass center of $\beta_l$ relative to the inertial reference frame, $\mathbf{D}_l$ as the position vector of the mass element $dm(\mathbf{D}_l)$ relative to the mass center of $\beta_l$, $\rho_g(\mathbf{D}_l)$ as the density of the mass element, where $dm(\mathbf{D}_l) = \rho_g(\mathbf{D}_l) dV(\mathbf{D}_l)$ and $dV(\mathbf{D}_l)$ is the volume element, $\mathbf{A}_l$ as the attitude matrix of the principal reference frame of $\beta_l$ relative to the inertial reference frame, $\mathbf{p}_l = m_l \dot{\mathbf{r}}_l$ as the linear momentum vector, $\mathbf{q}_l = \mathbf{A}_l \mathbf{D}_l + \mathbf{r}_l$ as the position vector of the mass element $dm(\mathbf{D}_l)$ relative to the inertial reference frame, $\mathbf{d}_l = \mathbf{A}_l \mathbf{D}_l$, $\mathbf{K}_l = \mathbf{r}_l \times \mathbf{p}_l + \mathbf{A}_l \mathbf{I}_l \Phi_l$ as the angular momentum vector of $\beta_l$, $\mathbf{G}_l = \mathbf{I}_l \Phi_l$, $\varepsilon_0$ as the permittivity of free space, $\mu_0$ as the permeability of vacuum, $\rho_{se}(\mathbf{D}_l)$ as the charge density of $\beta_l$, and $\mathbf{J}(\mathbf{D}_l)$ as the current density of $\beta_l$. For a vector $\mathbf{v} = [v_x, v_y, v_z]^\mathrm{T}$, we use $\hat{\mathbf{v}}$ to denote the matrix:

$$\widehat{\mathbf{v}} = \begin{pmatrix} 0 & -v_z & v_y \\ v_z & 0 & -v_x \\ -v_y & v_x & 0 \end{pmatrix}. \tag{1}$$

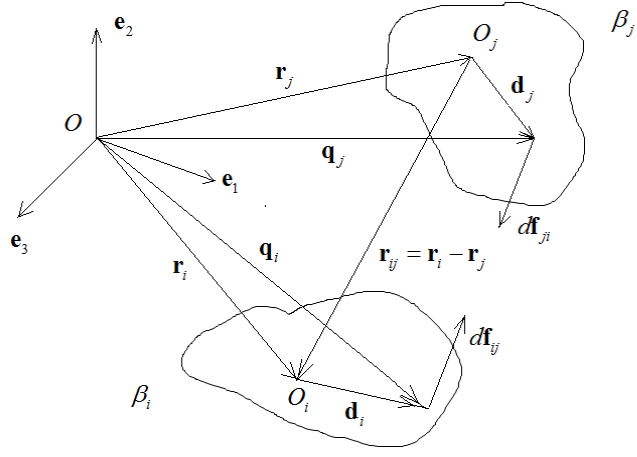

Fig. 1 The full N-body problem

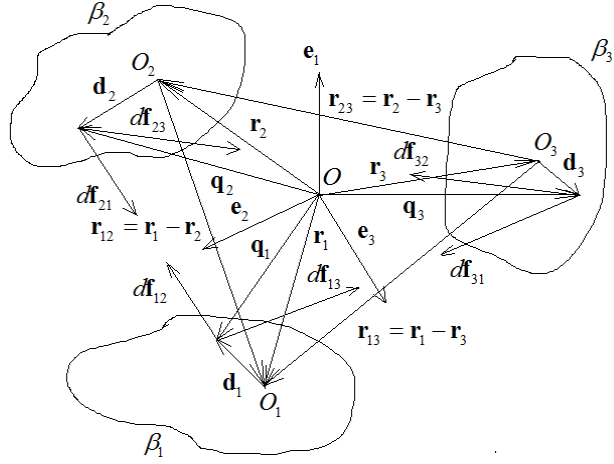

Fig. 2 The full 3-body problem

Then, the total gravitational potential energy of the system can be expressed as

$$U_g = -\sum_{k=1}^{n-1} \sum_{j=k+1}^{n} \int_{\beta_k} \int_{\beta_j} \frac{G\rho_g(\mathbf{D}_k)\rho_g(\mathbf{D}_j) dV(\mathbf{D}_j) dV(\mathbf{D}_k)}{\|\mathbf{q}_k - \mathbf{q}_j\|}$$

$$= -\sum_{k=1}^{n-1} \sum_{j=k+1}^{n} \int_{\beta_k} \int_{\beta_j} \frac{G\rho_g(\mathbf{D}_k)\rho_g(\mathbf{D}_j) dV(\mathbf{D}_j) dV(\mathbf{D}_k)}{\|\mathbf{A}_k \mathbf{D}_k - \mathbf{A}_j \mathbf{D}_j + \mathbf{r}_k - \mathbf{r}_j\|}. \tag{2}$$

The total kinetic energy of the system can be given by

$$T = \frac{1}{2}\sum_{k=1}^{n}\left(m_k \|\dot{\mathbf{r}}_k\|^2 + \langle \Phi_k, \mathbf{I}_k \Phi_k \rangle\right), \quad (3)$$

where $m_k$ is the mass of $\beta_k$. Thus, the total energy of the system is given as follows

$$H = T + U_g$$
$$= \frac{1}{2}\sum_{k=1}^{n}\left(m_k \|\dot{\mathbf{r}}_k\|^2 + \langle \Phi_k, \mathbf{I}_k \Phi_k \rangle\right) - \sum_{k=1}^{n-1}\sum_{j=k+1}^{n} \int_{\beta_k}\int_{\beta_j} G\rho_g(\mathbf{D}_k)\rho_g(\mathbf{D}_j)\frac{dV(\mathbf{D}_j)dV(\mathbf{D}_k)}{\|\mathbf{A}_k \mathbf{D}_k - \mathbf{A}_j \mathbf{D}_j + \mathbf{r}_k - \mathbf{r}_j\|}.$$

(4)

The total gravitational force acting on $\beta_k$ can be written as

$$\mathbf{f}_g^k = -G \sum_{j=1, j\neq k}^{n} \int_{\beta_k}\int_{\beta_j} \frac{(\mathbf{A}_k \mathbf{D}_k - \mathbf{A}_j \mathbf{D}_j + \mathbf{r}_k - \mathbf{r}_j)}{\|\mathbf{A}_k \mathbf{D}_k - \mathbf{A}_j \mathbf{D}_j + \mathbf{r}_k - \mathbf{r}_j\|^3} \rho_g(\mathbf{D}_k)\rho_g(\mathbf{D}_j) dV(\mathbf{D}_j) dV(\mathbf{D}_k). \quad (5)$$

Relative to the inertial space, the resultant gravitational torque acting on $\beta_k$ can be expressed as

$$\mathbf{n}_g^k = G \sum_{j=1, j\neq k}^{n} \int_{\beta_k}\int_{\beta_j} \frac{(\mathbf{A}_k \mathbf{D}_k + \mathbf{r}_k) \times (\mathbf{A}_j \mathbf{D}_j + \mathbf{r}_j)}{\|\mathbf{A}_k \mathbf{D}_k - \mathbf{A}_j \mathbf{D}_j + \mathbf{r}_k - \mathbf{r}_j\|^3} \rho_g(\mathbf{D}_k)\rho_g(\mathbf{D}_j) dV(\mathbf{D}_j) dV(\mathbf{D}_k). \quad (6)$$

Thus, the dynamical equation relative to the inertial space can be written in the following form:

$$\begin{cases} \dot{\mathbf{p}}_k = \mathbf{f}_g^k \\ \dot{\mathbf{r}}_k = \dfrac{\mathbf{p}_k}{m_k} \\ \dot{\mathbf{K}}_k = \mathbf{n}_g^k \\ \dot{\mathbf{A}}_k = \widehat{\boldsymbol{\psi}}_k \mathbf{A}_k \end{cases}, \quad k = 1, 2, \cdots n, \quad (7)$$

where $\boldsymbol{\psi}_k = \mathbf{A}_k \mathbf{I}_k^{-1} \mathbf{A}_k^{\mathrm{T}}(\mathbf{K}_k - \mathbf{r}_k \times \mathbf{p}_k)$, $\widehat{\boldsymbol{\psi}}_k$ is calculated with Eq. (1). The motion depends on the position vector $\mathbf{r}_k$, the linear momentum vector $\mathbf{p}_k = m_k \dot{\mathbf{r}}_k$, the angular momentum vector $\mathbf{K}_k$, and the attitude matrix $\mathbf{A}_k$. If we consider the gravitational potential energy, static electric potential energy, and the magnetic potential energy of the irregular bodies, the system includes the gravitational force, the static electric force, and the magnetic force, which makes this a full multi-body gravitoelectrodynamics system (Matthews and Hyde 2003; Fahnestock and Scheeres 2006; De Matos 2008). Appendix A presents the cases for gravitational potential energy, static electric potential energy, and the magnetic potential energy.

The dynamical equation has three conservative values, which are the first integrals

$$H = T + U_g, \quad \mathbf{p} = \sum_{k=1}^{n} \mathbf{p}_k, \quad \mathbf{K} = \sum_{k=1}^{n} \mathbf{K}_k .\qquad(8)$$

These three conservative values include the Jacobi integral, i.e., the energy integral, the linear momentum integral, and the angular momentum integral.

**3. Relative Motion and Equilibria**

Synchronous binary-asteroid systems, such as 809 Lundia, 854 Frostia, 1089 Tama, 4492 Debussy, and 1313 Berna, have been discovered in the solar system (Behrend et al. 2006; Descamps 2010; Jacobson and Scheeres 2011). These systems are cases of relative equilibrium where the orbits and attitudes of the respective bodies are gravitationally locked. The dwarf planet 134340 Pluto, as a trans-Neptunian object, has five known natural moonlets, Charon, Nix, Hydra, Kerberos, and Styx (Brozović et al. 2015). Charon was discovered in 1978, Nix and Hydra were discovered in 2005, Kerberos was discovered in 2011, and Styx was discovered in 2012. Pluto and Charon are gravitationally locked, and continuously present the same face to each other (Lee and Peale 2006; Buie et al. 2012). In addition, the orbit of Charon relative to Pluto is circular (Buie et al. 2012). To discuss the equilibrium of multiple celestial bodies, including synchronous binary-asteroid systems and gravitationally locked binaries in multiple body systems, we derive the relative motion equation, which is then employed to study the relative equilibria of systems comprised of multiple asteroids, the status of asteroids exhibiting partial gravitational locking, and the status of spin-orbit locked multiple-asteroid systems.

## 3.1 Relative Motion

In this section, the dynamical equation is expressed in the body-fixed frame of $\beta_n$ (see Fig. 3). We denote $\mathbf{r}_{kj} = \mathbf{r}_k - \mathbf{r}_j$, $\mathbf{R}_{kn} = \mathbf{A}_n^T \mathbf{r}_{kn} = \mathbf{A}_n^T (\mathbf{r}_k - \mathbf{r}_n)$, $\mathbf{P}_{kn} = \mathbf{A}_n^T \mathbf{p}_{kn}$ $= m_k \mathbf{A}_n^T (\dot{\mathbf{r}}_k - \dot{\mathbf{r}}_n)$, $\mathbf{A}_{kn} = \mathbf{A}_k^T \mathbf{A}_n$, $\mathbf{\Gamma}_{kn} = \mathbf{A}_{kn}^T \mathbf{G}_k$, $\mathbf{\Gamma}_n = \mathbf{G}_n$, $\mathbf{R}_{kj} = \mathbf{A}_j^T \mathbf{r}_{kj}$. Then,

$$\mathbf{G}_k = \mathbf{A}_k^T \mathbf{g}_k = \mathbf{I}_k \mathbf{\Phi}_k . \tag{9}$$

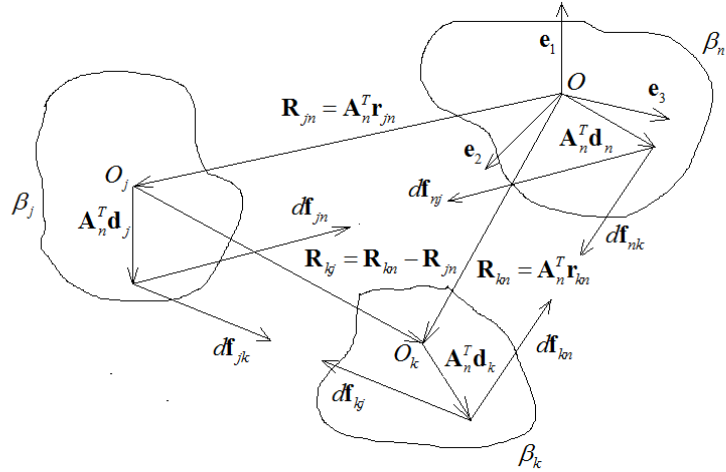

Fig. 3 The full N-body problem expressed in the body-fixed frame of $\beta_n$

Because $\mathbf{r}_{kn} = \mathbf{r}_k - \mathbf{r}_n$ and $\mathbf{r}_{kn} = \mathbf{r}_k - \mathbf{r}_n$, we have

$$\mathbf{r}_{kj} = \mathbf{r}_{kn} - \mathbf{r}_{jn} . \tag{10}$$

In addition,

$$\mathbf{R}_{kn} - \mathbf{R}_{jn} = \mathbf{A}_n^T \mathbf{r}_{kn} - \mathbf{A}_n^T \mathbf{r}_{jn} = \mathbf{A}_n^T \mathbf{r}_{kj} = \mathbf{A}_n^T (\mathbf{r}_k - \mathbf{r}_j). \tag{11}$$

Then, the total gravitational potential energy can be written relative to the body-fixed frame of $\beta_n$ as follows:

$$U_g = -\sum_{k=1}^{n-1} \sum_{j=k+1}^{n} \int_{\beta_k} \int_{\beta_j} \frac{G \rho_g(\mathbf{D}_k) \rho_g(\mathbf{D}_j) dV(\mathbf{D}_j) dV(\mathbf{D}_k)}{\|\mathbf{A}_n^T \mathbf{A}_k \mathbf{D}_k - \mathbf{A}_n^T \mathbf{A}_j \mathbf{D}_j + \mathbf{R}_{kn} - \mathbf{R}_{jn}\|}. \tag{12}$$

If we denote $U = U_g$, the dynamical equation of the system can be given by

$$\begin{cases} \dot{\mathbf{P}}_{kn} = \mathbf{P}_{kn} \times \Phi_n - \dfrac{\partial U}{\partial \mathbf{R}_{kn}} \\[4pt] \dot{\mathbf{R}}_{kn} = \mathbf{R}_{kn} \times \Phi_n + \dfrac{\mathbf{P}_{kn}}{m_k} \\[4pt] \dot{\Gamma}_k = \Gamma_k \times \Phi_n + \boldsymbol{\mu}_k, \quad \dot{\Gamma}_n = \Gamma_n \times \Phi_n + \boldsymbol{\mu}_n \\[4pt] \dot{\mathbf{A}}_{kn} = \mathbf{A}_{kn} \hat{\Phi}_n - \hat{\Phi}_k \mathbf{A}_{kn}, \quad \dot{\mathbf{A}}_n = \mathbf{A}_n \hat{\Phi}_n \end{cases}, \tag{13}$$

where $k = 1, 2, \cdots, n-1$, $\Phi_n = \mathbf{I}_n^{-1} \Gamma_n$, and $\Phi_k = \mathbf{I}_k^{-1} \mathbf{A}_{kn} \Gamma_k$.

The torque is

$$\begin{cases} \boldsymbol{\mu}^k = \boldsymbol{\mu}_g^k, \quad k = 1, 2, \cdots, n-1 \\ \boldsymbol{\mu}^n = \boldsymbol{\mu}_g^n \end{cases}, \tag{14}$$

where

$$\boldsymbol{\mu}_g^k = -G \sum_{k=1}^{n-1} \sum_{j=k+1}^{n} \int_{\beta_k} \int_{\beta_j} \mathbf{A}_n^T \mathbf{A}_k \mathbf{D}_k \times \frac{\left(\mathbf{A}_n^T \mathbf{A}_k \mathbf{D}_k - \mathbf{A}_n^T \mathbf{A}_j \mathbf{D}_j + \mathbf{R}_{kn} - \mathbf{R}_{jn}\right)}{\|\mathbf{A}_n^T \mathbf{A}_k \mathbf{D}_k - \mathbf{A}_n^T \mathbf{A}_j \mathbf{D}_j + \mathbf{R}_{kn} - \mathbf{R}_{jn}\|^3} \rho_g(\mathbf{D}_k) \rho_g(\mathbf{D}_j) dV(\mathbf{D}_j) dV(\mathbf{D}_k)$$

$$\tag{15}$$

and

$$\boldsymbol{\mu}_g^n = G \sum_{k=1}^{n-1} \sum_{j=k+1}^{n} \int_{\beta_k} \int_{\beta_j} \mathbf{D}_n \times \frac{\left(\mathbf{A}_n^T \mathbf{A}_k \mathbf{D}_k - \mathbf{D}_n + \mathbf{R}_{kn}\right)}{\|\mathbf{A}_n^T \mathbf{A}_k \mathbf{D}_k - \mathbf{D}_n + \mathbf{R}_{kn}\|^3} \rho_g(\mathbf{D}_k) \rho_g(\mathbf{D}_j) dV(\mathbf{D}_j) dV(\mathbf{D}_k). \tag{16}$$

For details about the total static electric potential energy and the total magnetic potential energy, see Appendix B.

## 3.2 Relative Equilibria

From Eq. (13), we know that a relative equilibrium satisfies the following equation:

$$\begin{cases} \mathbf{f}_{\mathbf{P}_k} \triangleq \mathbf{P}_{kn} \times \Phi_n - \dfrac{\partial U}{\partial \mathbf{R}_{kn}} = 0 \\ \mathbf{f}_{\mathbf{R}_k} \triangleq \mathbf{R}_{kn} \times \Phi_n + \dfrac{\mathbf{P}_{kn}}{m_k} = 0 \\ \mathbf{f}_{\mathbf{A}_k} \triangleq \mathbf{A}_{kn}\widehat{\Phi}_n - \widehat{\Phi}_k \mathbf{A}_{in} = 0 \\ \mathbf{f}_{\Gamma_k} \triangleq \Gamma_k \times \Phi_n + \boldsymbol{\mu}_k = 0 \\ \mathbf{f}_{\Gamma_n} \triangleq \Gamma_n \times \Phi_n + \boldsymbol{\mu}_n = 0 \end{cases}, \qquad (17)$$

where $k = 1, 2, \cdots, n-1$. Because the equilibrium is relative, $\dot{\mathbf{A}}_n = \mathbf{A}_n \widehat{\Phi}_n = 0$ is naturally established. A total of $12n-9$ equations are represented in Eq. (17). Thus, $12n-9$ eigenvalues exist for the state of relative equilibrium.

We denote

$$\begin{cases} \mathbf{X} = [\mathbf{P}_1, \cdots, \mathbf{P}_{n-1}, \mathbf{R}_1, \cdots, \mathbf{R}_{n-1}, \mathbf{A}_1, \cdots, \mathbf{A}_{n-1}, \Gamma_1, \cdots, \Gamma_n] \\ \mathbf{f} = [\mathbf{f}_{\mathbf{P}_1}, \cdots, \mathbf{f}_{\mathbf{P}_{n-1}}, \mathbf{f}_{\mathbf{R}_1}, \cdots, \mathbf{f}_{\mathbf{R}_{n-1}}, \mathbf{f}_{\mathbf{A}_1}, \cdots, \mathbf{f}_{\mathbf{A}_{n-1}}, \mathbf{f}_{\Gamma_1}, \cdots, \mathbf{f}_{\Gamma_n}] \end{cases}, \qquad (18)$$

$$\begin{cases} \delta \mathbf{P}_k = \mathbf{P}_k(t) - \mathbf{P}_k(t_0), (k = 1, \cdots, n-1) \\ \delta \mathbf{R}_k = \mathbf{R}_k(t) - \mathbf{R}_k(t_0), (k = 1, \cdots, n-1) \\ \delta \mathbf{A}_k = \mathbf{A}_k(t) - \mathbf{A}_k(t_0), (k = 1, \cdots, n-1) \\ \delta \Gamma_j = \Gamma_j(t) - \Gamma_j(t_0), (j = 1, \cdots, n) \end{cases}, \qquad (19)$$

$$\nabla \mathbf{f}_{\mathbf{P}_k} = \dfrac{\partial \mathbf{f}_{\mathbf{P}_k}}{\partial \mathbf{X}}, \nabla \mathbf{f}_{\mathbf{R}_k} = \dfrac{\partial \mathbf{f}_{\mathbf{R}_k}}{\partial \mathbf{X}}, \nabla \mathbf{f}_{\mathbf{A}_k} = \dfrac{\partial \mathbf{f}_{\mathbf{A}_k}}{\partial \mathbf{X}}, \nabla \mathbf{f}_{\Gamma_j} = \dfrac{\partial \mathbf{f}_{\Gamma_j}}{\partial \mathbf{X}}, (k = 1, \cdots, n-1; j = 1, \cdots, n), \qquad (20)$$

and

$$\nabla \mathbf{f} = [\nabla \mathbf{f}_{\mathbf{P}_1}, \cdots, \nabla \mathbf{f}_{\mathbf{P}_{n-1}}, \nabla \mathbf{f}_{\mathbf{R}_1}, \cdots, \nabla \mathbf{f}_{\mathbf{R}_{n-1}}, \nabla \mathbf{f}_{\mathbf{A}_1}, \cdots, \nabla \mathbf{f}_{\mathbf{A}_{n-1}}, \nabla \mathbf{f}_{\Gamma_1}, \cdots, \nabla \mathbf{f}_{\Gamma_n}]. \qquad (21)$$

Thus, the eigenvalues of $\nabla \mathbf{f}$ are constituted of the eigenvalues of $\nabla \mathbf{f}_{\mathbf{P}_k}, \nabla \mathbf{f}_{\mathbf{R}_k}, \nabla \mathbf{f}_{\mathbf{A}_k},$ and $\nabla \mathbf{f}_{\Gamma_j} \ (k = 1, \cdots, n-1; j = 1, \cdots, n)$, and we denote eigenvalues of $\nabla \mathbf{f}$ as $\lambda_1, \cdots, \lambda_k \ (k = 12n-9)$. Dynamical equation (17) can be linearized under the condition of relative equilibrium as

$$\begin{cases} \delta \dot{\mathbf{P}}_k = \nabla \mathbf{f}_{\mathbf{P}_k} \cdot \delta \mathbf{P}_k, (k=1,\cdots,n-1) \\ \delta \dot{\mathbf{R}}_k = \nabla \mathbf{f}_{\mathbf{R}_k} \cdot \delta \mathbf{R}_k, (k=1,\cdots,n-1) \\ \delta \dot{\mathbf{A}}_k = \nabla \mathbf{f}_{\mathbf{A}_k} \cdot \delta \mathbf{A}_k, (k=1,\cdots,n-1) \\ \delta \dot{\mathbf{\Gamma}}_j = \nabla \mathbf{f}_{\mathbf{\Gamma}_j} \cdot \delta \mathbf{\Gamma}_j, (j=1,\cdots,n) \end{cases} \quad (22)$$

We denote the number of completely relative equilibria as $J$ and the $j$-th relative equilibria as $E_j$. **f** Topological degree theory (Mawhin and Willem, 1989) in conjunction with the function provides

$$\sum_{j=1}^{J} \left[ \operatorname{sgn} \prod_{k=1}^{12n-9} \lambda_k (E_j) \right] = \sum_{j=1}^{J} \left[ \operatorname{sgn} \det (\nabla \mathbf{f}) \right] = \deg(\mathbf{f}, \ \Xi, \ 0) = \text{const}, \quad (23)$$

where $\Xi$ is the definition domain of $\mathbf{f}$, $\deg(\cdot)$ is the topological degree, and $\operatorname{sgn}(\cdot)$ is the symbolic function. A non-degenerate equilibrium requires all the eigenvalues to be non-zero. From Eq. (23), one can deduce that the number of non-degenerate equilibria for a system of multiple irregular-shaped bodies varies in pairs. When the system parameters vary, it is impossible for a degenerate equilibrium to change to an odd number of non-degenerate equilibria. In fact, the degenerate equilibrium has four choices: i) annihilate each other; ii) disappear and produce $\alpha(\alpha \in Z, \alpha > 0)$ degenerate equilibria, where $Z$ represents the integer set; iii) disappear and produce $2\beta(\beta \in Z, \beta > 0)$ non-degenerate equilibria; iv) disappear and produce $\alpha(\alpha \in Z, \alpha > 0)$ degenerate equilibria and $2\beta(\beta \in Z, \beta > 0)$ non-degenerate equilibria. For an equilibrium $E_j$, we denote $\operatorname{ind}(E_j) \triangleq \operatorname{sgn} \prod_{k=1}^{12n-9} \lambda_k (E_j)$. Then, if the equilibrium $E_j$ is degenerate, we have $\operatorname{ind}(E_j) = 0$; otherwise, $\operatorname{ind}(E_j) = 1$.

If there are only two bodies, i.e. $\beta_1$ and $\beta_2$, and $\beta_1$ is a massless body, then the relative equilibria of the system are equilibrium points for $\beta_1$ in the body-fixed frame of $\beta_2$. More details in similar types of problems about the stability, topological classifications, and bifurcations of relative equilibria can be found in Jiang et al. (2015a, 2015b, 2015c, 2016).

**3.3 Partial Gravitational Locking of Multiple Irregular Celestial Bodies**

A number of multiple celestial body systems that exhibit partial gravitational locking are known to exist, such as 134340 Pluto (Brozović et al. 2015; Showalter and Hamilton 2015). As already discussed, Pluto and Charon are gravitationally locked; however, the other moonlets of the system, are not gravitationally locked (Brozović et al. 2015), leading to a partially-locked system. Suppose that $n_1$ $(n_1 < n)$ asteroids are gravitationally locked to each other in a system, the $n$-th asteroid $\beta_n$ is arbitrarily selected from one of the $n_1$ locked asteroids, and all other locked asteroids' are denoted by subscripts $1, 2, ..., n_1 - 1$. Then, the relative equilibrium satisfies the following relations.

$$\begin{cases} \mathbf{f}_{\mathbf{P}_k} \triangleq \mathbf{P}_{kn} \times \Phi_n - \dfrac{\partial U}{\partial \mathbf{R}_{kn}} = 0 \\ \mathbf{f}_{\mathbf{R}_k} \triangleq \mathbf{R}_{kn} \times \Phi_n + \dfrac{\mathbf{P}_{kn}}{m_k} = 0 \\ \mathbf{f}_{\mathbf{A}_k} \triangleq \mathbf{A}_{kn} \widehat{\Phi}_n - \widehat{\Phi}_k \mathbf{A}_{kn} = 0 \\ \mathbf{f}_{\Gamma_k} \triangleq \Gamma_k \times \Phi_n + \boldsymbol{\mu}_k = 0 \\ \mathbf{f}_{\Gamma_n} \triangleq \Gamma_n \times \Phi_n + \boldsymbol{\mu}_n = 0 \end{cases} \quad k = 1, 2, \cdots, n_1 - 1. \quad (24)$$

In this case, there are $12n_1 - 9$ equations in Eq. (24), and, thus, $12n_1 - 9$ eigenvalues for the relative equilibrium. The state of partial gravitational locking is represented by motion around a stable relative equilibrium, indicative of the cycloidal motion of locked asteroids.

The corresponding formula of Eq. (24) is $\sum_{j=1}^{J}\left[\text{sgn}\prod_{k=1}^{12n_1-9}\lambda_k(E_j)\right]=const$. We can also deduce that the number of non-degenerate configurations exhibiting partial gravitational locking for these $n_1$ asteroids in the system varies in pairs. If the system parameters vary (for example, the mass, shape, or the mass distribution of at least one asteroid changes), the degenerate partially-locked configuration again has four choices. In addition, for a stable configuration of multiple irregular celestial bodies in a partially-locked state, the orbits and attitudes of the locked bodies are not constant, and their orbits and attitudes exhibit periodic or quasi-periodic variation. Moreover, stable configurations occur for systems comprised of bodies of suitable sizes and distances of separation.

**3.4 Spin-Orbit Locked Multiple-asteroid Systems**

In this section, we consider spin-orbit locked multiple-asteroid systems, where one or more asteroids exhibit synchronous axial rotation and revolution. As an example, the near-Earth triple-asteroid system 136617 1994 CC is comprised of two moonlets (Brozović et al. 2011), where only the inner moonlet is spin-orbit locked. We assume that the *n*-th asteroid $\beta_n$ is the largest body in the system, asteroids reside in spin-orbit locked states, there are $n_2$ $(n_2<n)$ asteroids belonging to the spin-orbit locked cases, and the subscripts of these spin-orbit locked asteroids are denoted as 1, 2, ..., $n_2$. Then, we have the following expressions.

$$\begin{cases} \mathbf{f}_{\mathbf{P}_k} \triangleq \mathbf{P}_{kn} \times \Phi_n - \dfrac{\partial U}{\partial \mathbf{R}_{kn}} = 0 \\ \mathbf{f}_{\mathbf{R}_k} \triangleq \mathbf{R}_{kn} \times \Phi_n + \dfrac{\mathbf{P}_{kn}}{m_k} = 0 \qquad k=1,\ 2;\cdot\ n_2. \\ \mathbf{f}_{\mathbf{A}_k} \triangleq \mathbf{A}_{kn}\hat{\Phi}_n - \hat{\Phi}_k \mathbf{A}_{kn} = 0 \\ \mathbf{f}_{\mathbf{\Gamma}_k} \triangleq \mathbf{\Gamma}_k \times \Phi_n + \mathbf{\mu}_k = 0 \end{cases} \tag{25}$$

In this case, there are $12n_2$ equations in Eq. (25), and, therefore, $12n_2$ eigenvalues for the spin-orbit locked status. As such, for a triple-asteroid system with a single moonlet in a spin-orbit locked state, there are 12 equations for the spin-orbit locked status. The spin-orbit locked status of an asteroid reflects motion around a stable relative equilibrium, indicating a cycloidal motion for spin-orbit locked asteroids.

**3.5 Nonlinear Changes in Orbit Parameters for High Size Ratio Triple-Asteroid Systems**

For high size ratio triple-asteroid systems, nonlinear changes in orbit parameters can be estimated theoretically. We denote $a$ as the length of the semi-major axis, $e$ as the eccentricity, $i$ as the inclination, $\Omega$ as the longitude of the ascending node, $\omega$ as the argument of periapsis, $M$ as the mean anomaly, $m$ as the mass, $p = a(1-e^2)$, $\eta = \sqrt{\dfrac{Gm}{a^3}}$, $J_2$ as the second order coefficient of the nonspherical gravitational potential of the primary body, and $R$ as the primary body's mean radius. Let subscripts M1 and M2 represent the orbital parameters of Moonlet 1 and Moonlet 2. Then, the rate of change (Kozai 1959) of $\Omega$ for Moonlet $k$ ($k$ = 1, 2) caused by the primary body's nonspherical gravitational potential can be expressed as

$$\frac{d\Omega_{Mk}}{dt} = -\frac{3}{2} J_2 \left(\frac{R}{p_{Mk}}\right)^2 \eta_{Mk} \cos i_{Mk}. \tag{26}$$

The above equation indicates that the rate of change of $\Omega$ varies for different moonlets, which is related to $a$ and $e$ (through $p_{Mk}$) as well as $i_{Mk}$. When $i_{M1} = i_{M2}$, the rate of change of $\Omega$ for the inner moonlet is greater than that of the outer moonlet. Therefore, the ascending node of the inner moonlet moves faster than the outer moonlet.

Using the motion equation of inclination (Sidi 1997):

$$\frac{di}{dt} = -\frac{1}{i\eta a^2} \frac{\partial U_p}{\partial \Omega}, \tag{27}$$

one can easily obtain the rate of change of $i$ for Moonlet $k$ ($k = 1, 2$) caused by the primary body's nonspherical gravitational potential:

$$\begin{cases} \frac{di_{M1}}{dt} = \frac{3}{8} n_{M1} \sigma_{M1} \left(\frac{\eta_{M2}}{\eta_{M1}}\right)^2 \sin(\Omega_{M1} - \Omega_{M2}) \left[2\cos(\Omega_{M1} - \Omega_{M2}) \sin i_{M1} + \sin 2i_{M2} \cos i_{M1} \right. \\ \qquad \left. -2\cos^2 i_{M2} \sin i_{M1} \cos(\Omega_{M1} - \Omega_{M2})\right] \\ \frac{di_{M2}}{dt} = \frac{3}{8} n_{M2} \sigma_{M2} \left(\frac{\eta_{M1}}{\eta_{M2}}\right)^2 \sin(\Omega_{M2} - \Omega_{M1}) \left[2\cos(\Omega_{M2} - \Omega_{M1}) \sin i_{M2} + \sin 2i_{M1} \cos i_{M2} \right. \\ \qquad \left. -2\cos^2 i_{M1} \sin i_{M2} \cos(\Omega_{M2} - \Omega_{M1})\right] \end{cases}, \tag{28}$$

where $U_p$ is the perturbing potential function, $m_{primary}$ is the primary body's mass, $\sigma_{M1} = \frac{m_{M1}}{m_{M1} + m_{primary}}$, and $\sigma_{M2} = \frac{m_{M2}}{m_{M2} + m_{primary}}$. Equation (28) indicates that the two moonlets in a high size ratio triple-asteroid system have changing inclinations, although the primary body's nonspherical gravitational potential has no long-term influence on the two moonlets' orbital elements. Therefore, the changing inclinations of the two moonlets are caused by their gravitational effect on each other.

# 4. Dynamical Configurations for Five Triple-asteroid Systems and a Six-body System

In this section, we present the orbital and attitude motion of several realistic multiple-asteroid systems through simulations of the full N rigid body problem. The dynamical configurations are calculated by Eq. (7). Only the gravitational potential is considered. The configurations of the triple-asteroid systems 45 Eugenia, 87 Sylvia, 93 Minerva, 216 Kleopatra, and 136617 1994CC, and the six-body system 134340 Pluto are computed.

## 4.1 Numerical Method

To study the coupled orbital and rotational dynamics for a multiple-asteroid system, a complete simulation must include the mutual gravitational potential between the components and the perturbations arising from the irregular shape of the primary body. In this paper, a high resolution model of point masses on a dense hexagonal grid is applied to fill the body, as shown in Fig. 4. From the perspective of asteroid structure, this model can generally reflect the loose and porous characteristics of the majority of asteroids between ~100 m and ~100 km in size (Richardson et al. 2002), which has been successfully used to study the collision process and aggregation dynamics in space (e.g., Matthews and Hyde 2004; Michel and Richardson 2013). Furthermore, because the mutual gravitational force and torque can be explicitly computed as summations over the point masses in different bodies, this method can easily extend the number of components. Richardson (1995) has given a detailed introduction to this method. In brief, the dynamical behavior of each body can be described by the equations of motion of rigid-bodies. The motion of the mass center of a body obeys Newton's equations, and is integrated using the leapfrog algorithm, while the rotation is derived from the rigid-body Euler equations, and is solved through a high-order time-adaptive Runge-Kutta 7/8 method. The orientation of the aggregate is described by the four quaternions. The integration time-span is chosen as 0.1s, which can make the conservation of energy and angular momentum unchanged for the system. The accuracy of the mixed method can be seen in the figure of the relative energy and relative angular momentum.

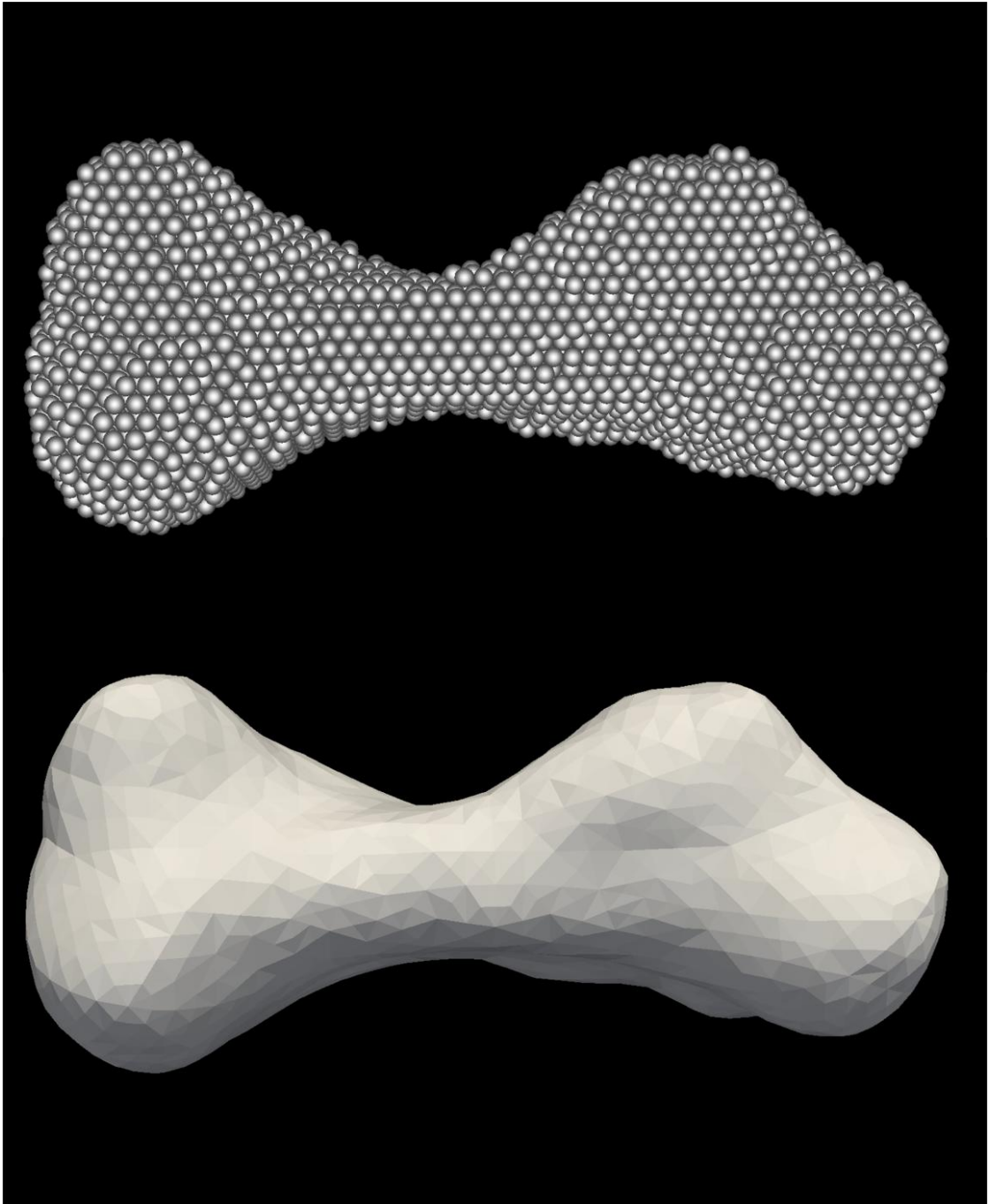

Fig. 4 An illustration of the use of a high resolution model comprised of point masses to fill an irregular celestial body (the primary body of the triple-asteroid system 216 Kleopatra was employed as an example).

## 4.2 Initial Conditions

The five triple-asteroid systems considered represent those systems out of nine discovered systems for which orbital data exist for the moonlets. The triple-asteroid systems can be arranged in order of size from large to small based on the diameters of the primary bodies as follows: 87 Sylvia (Fang et al. 2012; Berthier et al. 2014) with a size of 286 km; 45 Eugenia (Marchis et al. 2010; Beauvalet et al. 2011; Beauvalet and Marchis 2014) with a size of 206.14 km; 93 Minerva (Marchis et al. 2011; Marchis et al. 2013) with a size of 141.6 km; 216 Kleopatra (Descamps et al. 2010) with a size of 135 km; and 136617 1994 CC (Brozović et al. 2010, 2011; Fang et al. 2011) with a size of 0.62 km. Among these large-size triple-asteroid systems, only 45 Eugenia, 87 Sylvia, 93 Minerva, and 216 Kleopatra conform to a radar shape model (Neese 2004). For 45 Eugenia, 87 Sylvia, 93 Minerva, and 216 Kleopatra, the model includes the irregular shape of the major body, and sphere shape of two moonlets. For 136617 1994CC, the model includes the ellipsoid shape of the major body, and sphere shape of two moonlets. The three axis length of major body for 136617 1994CC is $0.69 \times 0.67 \times 0.64$ km (Brozović et al. 2011). For the six-body system 134340 Pluto, the model includes the ellipsoid shape of Pluto and Charon, and sphere shape of other four moonlets. The three axis radius of Pluto and Charon are assumed to be $1193 \times 1180 \times 1169$ km and $605 \times 603.6 \times 602.2$ km, respectively.

Table 1 and Table 2 in the Appendix B list the initial conditions applied for calculation of the dynamical configurations. From Table 1, we see that 45 Eugenia, 87 Sylvia, 93 Minerva, and 216 Kleopatra are high size ratio triple-asteroid systems, while 136,617 1994CC is not.

## 4.3 Calculation of Dynamical Configurations

Figs. 5 to 14 show the dynamical calculations of the five triple-asteroid systems, and Figs. 15 and 17 show the dynamical calculations of the six-body system. The orbit parameters of the moonlets of the triple-asteroid systems are expressed in terms of the centroid coordinates of the primary body, where the z axis is coincident with the primary's spin axis while the xy plane is coincident with the primary body's equator. For 134340 Pluto, the orbit parameters of Charon are expressed in terms of Pluto's centroid coordinates while the orbit parameters of the other four moonlets are expressed in terms of the centroid inertial frame of the six-body system.

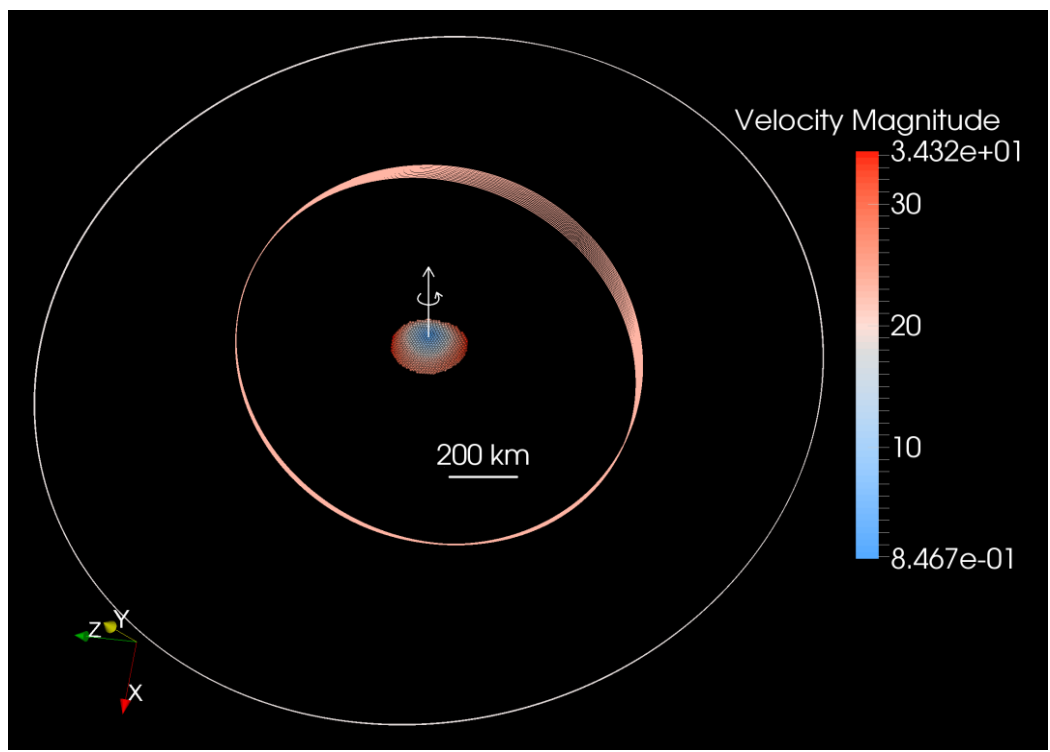

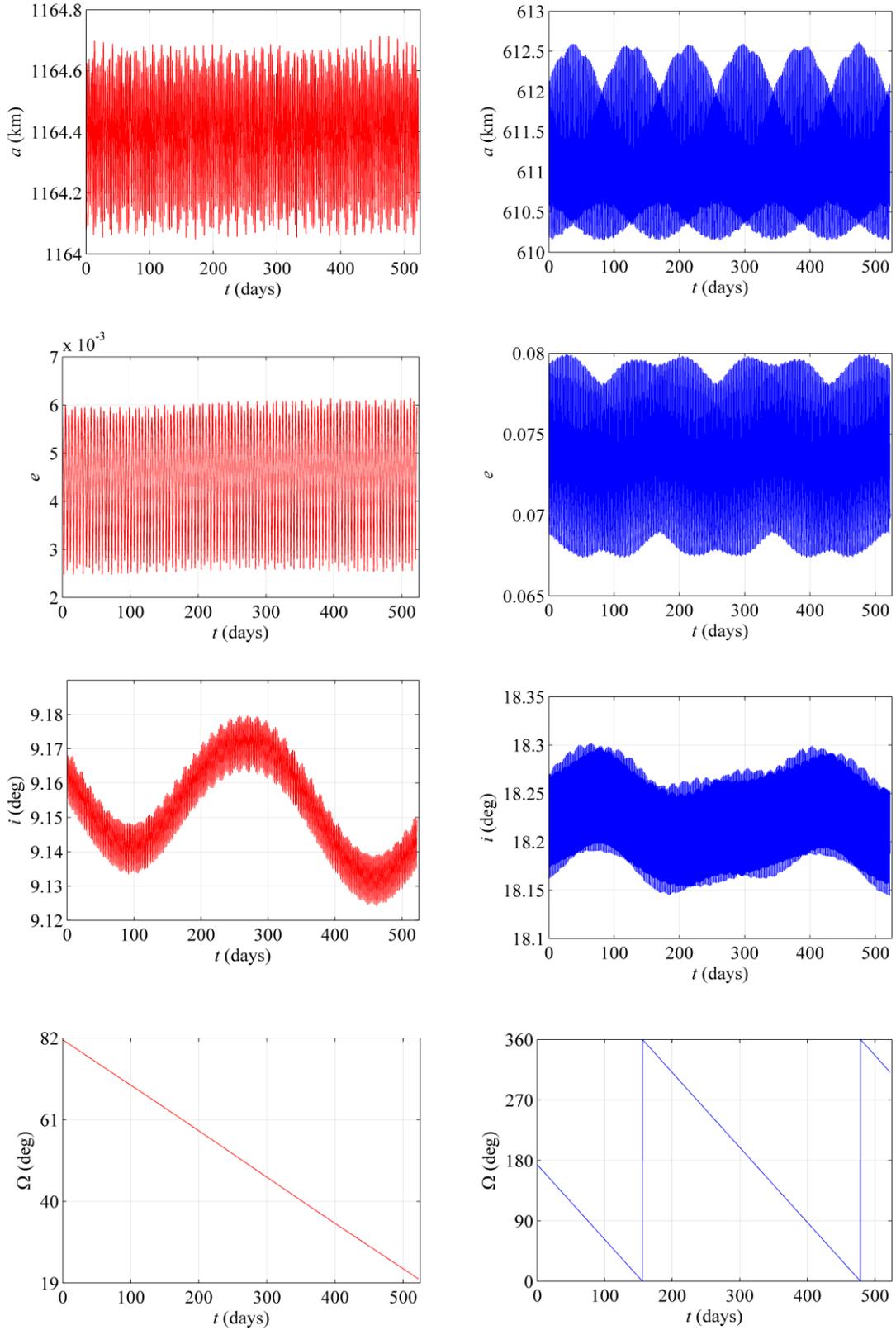

Fig. 5 Dynamical calculation of the triple-asteroid system 45 Eugenia: the dark background figure shows the orbits of the two moonlets Princesse and Petit-Prince, and the colorbar indicates the speed; the two columns show the semi-major axis, eccentricity, inclination, and the longitude of the ascending node for Princesse (left) and Petit-Prince (right).

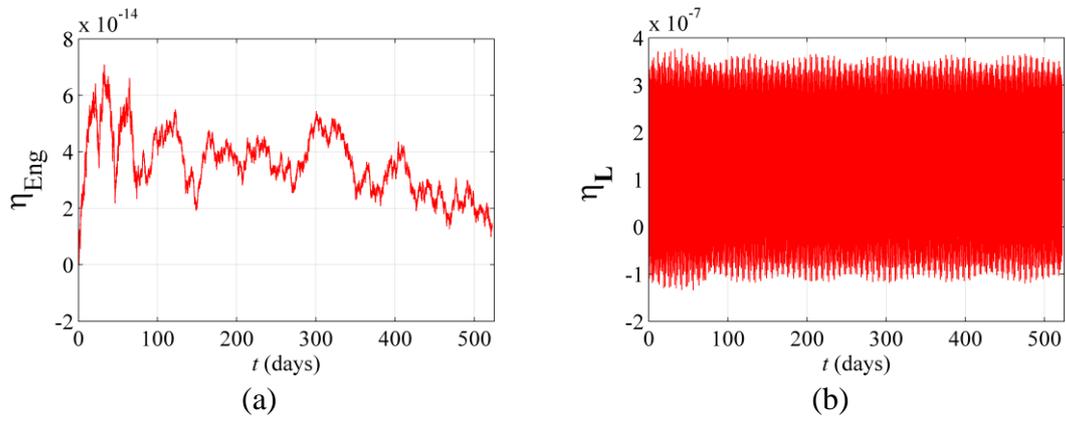

Fig. 6 The error of the energy and the error of the angular momentum for the calculation of the triple-asteroid system 45 Eugenia. (a) shows the relative energy while (b) shows the relative angular momentum.

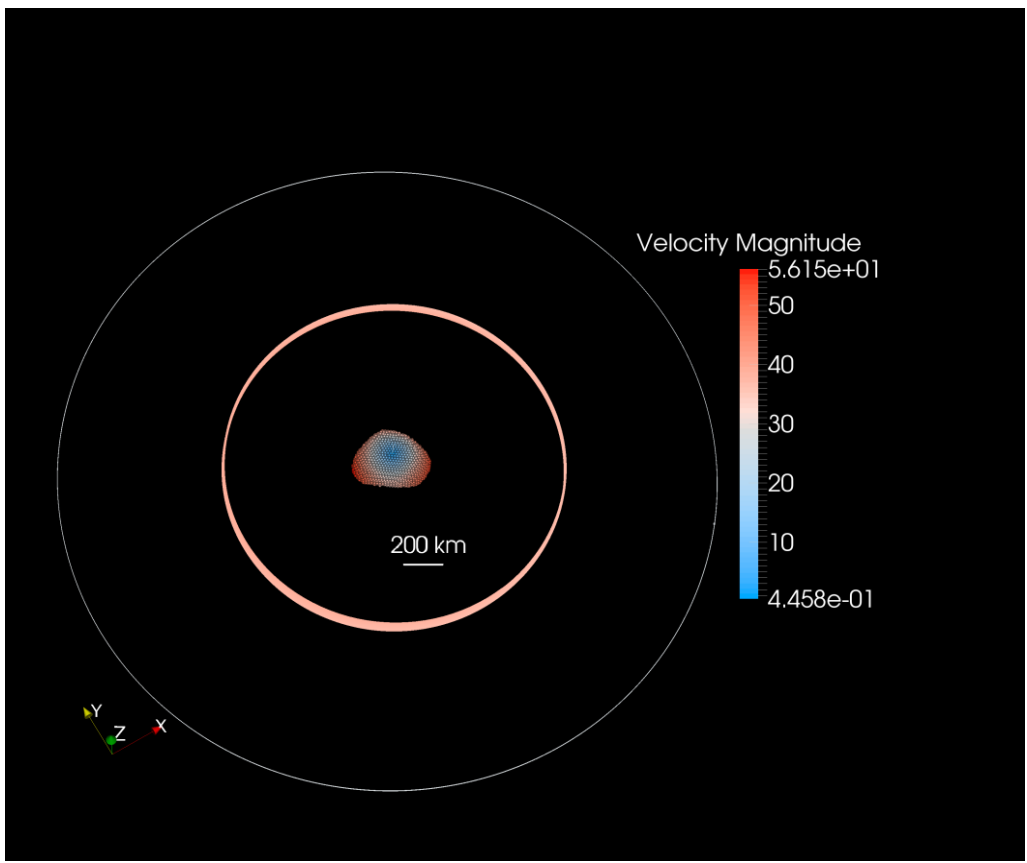

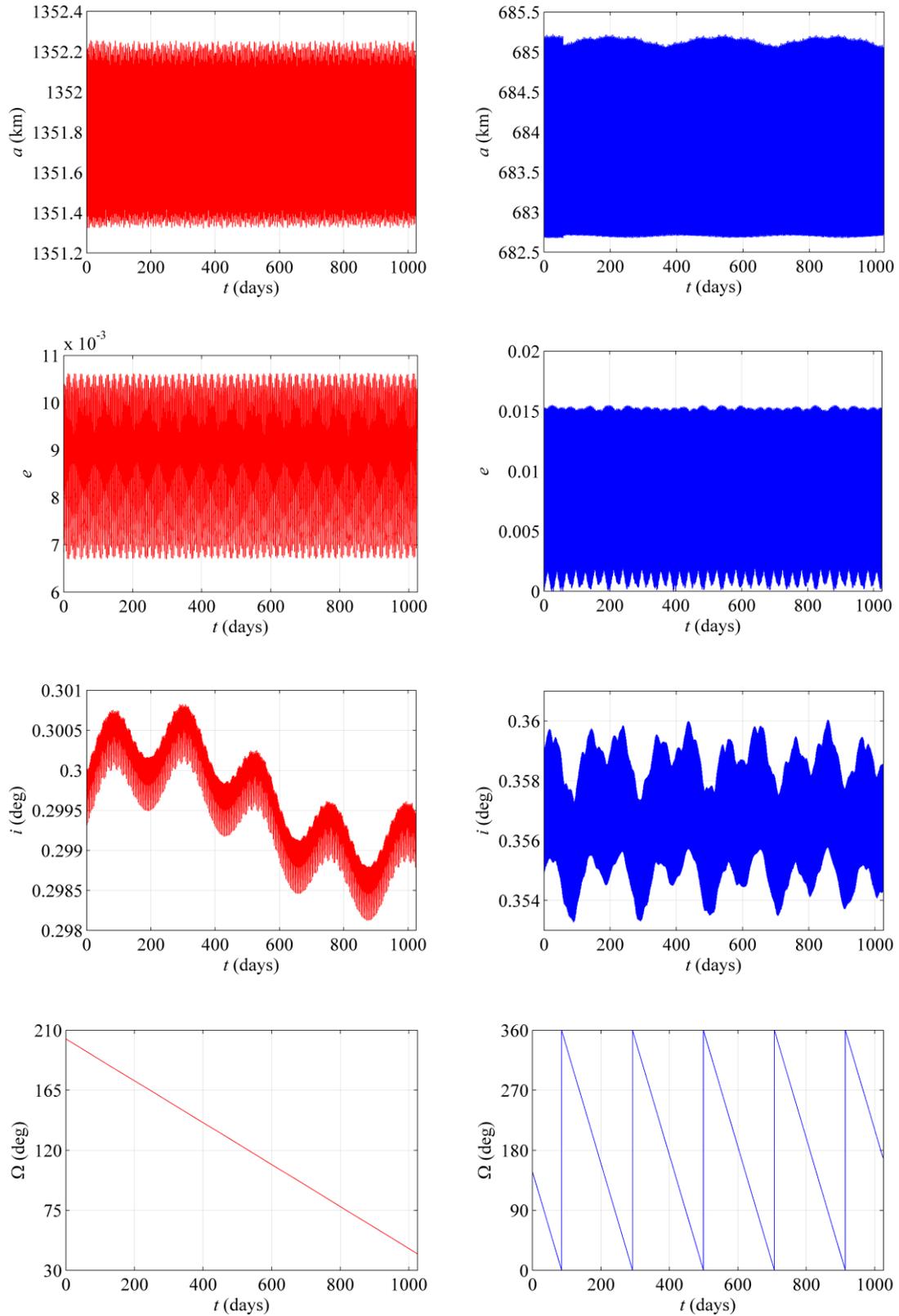

Fig. 7 Dynamical calculation of the triple-asteroid system 87 Sylvia: the dark background figure shows the orbits of the two moonlets Romulus and Remus, and the colorbar indicates the speed; the two columns show the semi-major axis, eccentricity, inclination, and the longitude of the ascending node for Romulus (left) and Remus (right).

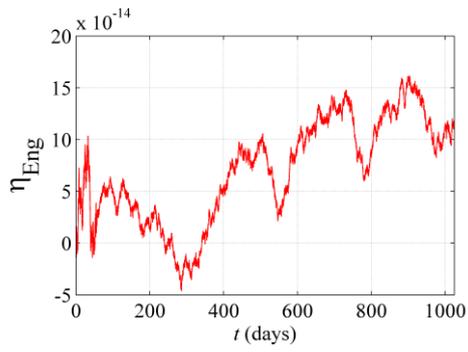
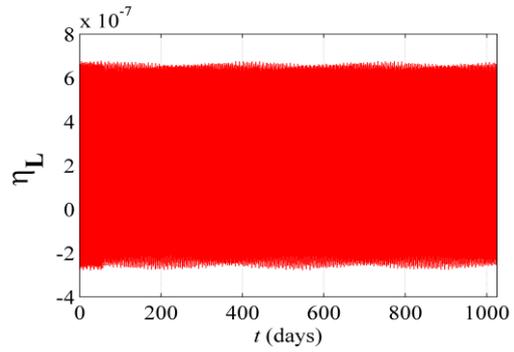

(a)                                            (b)

Fig. 8 The error of the energy and the error of the angular momentum for the calculation of the triple-asteroid system 87 Sylvia. (a) shows the relative energy while (b) shows the relative angular momentum.

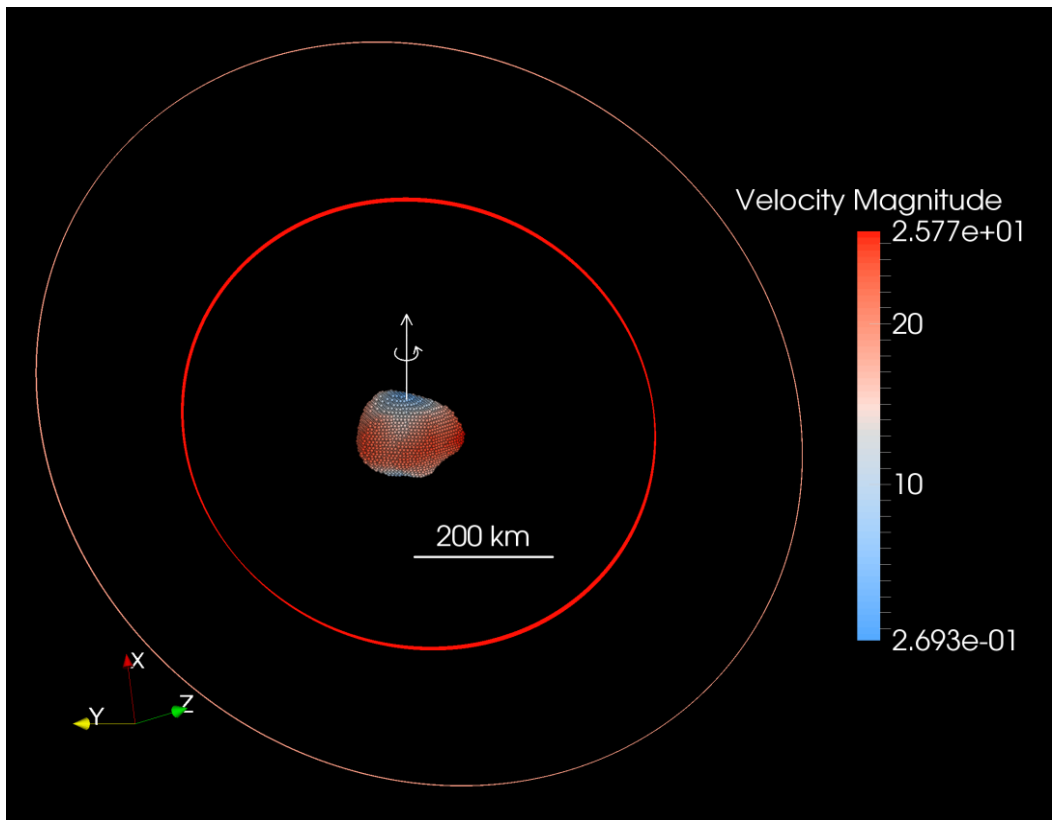

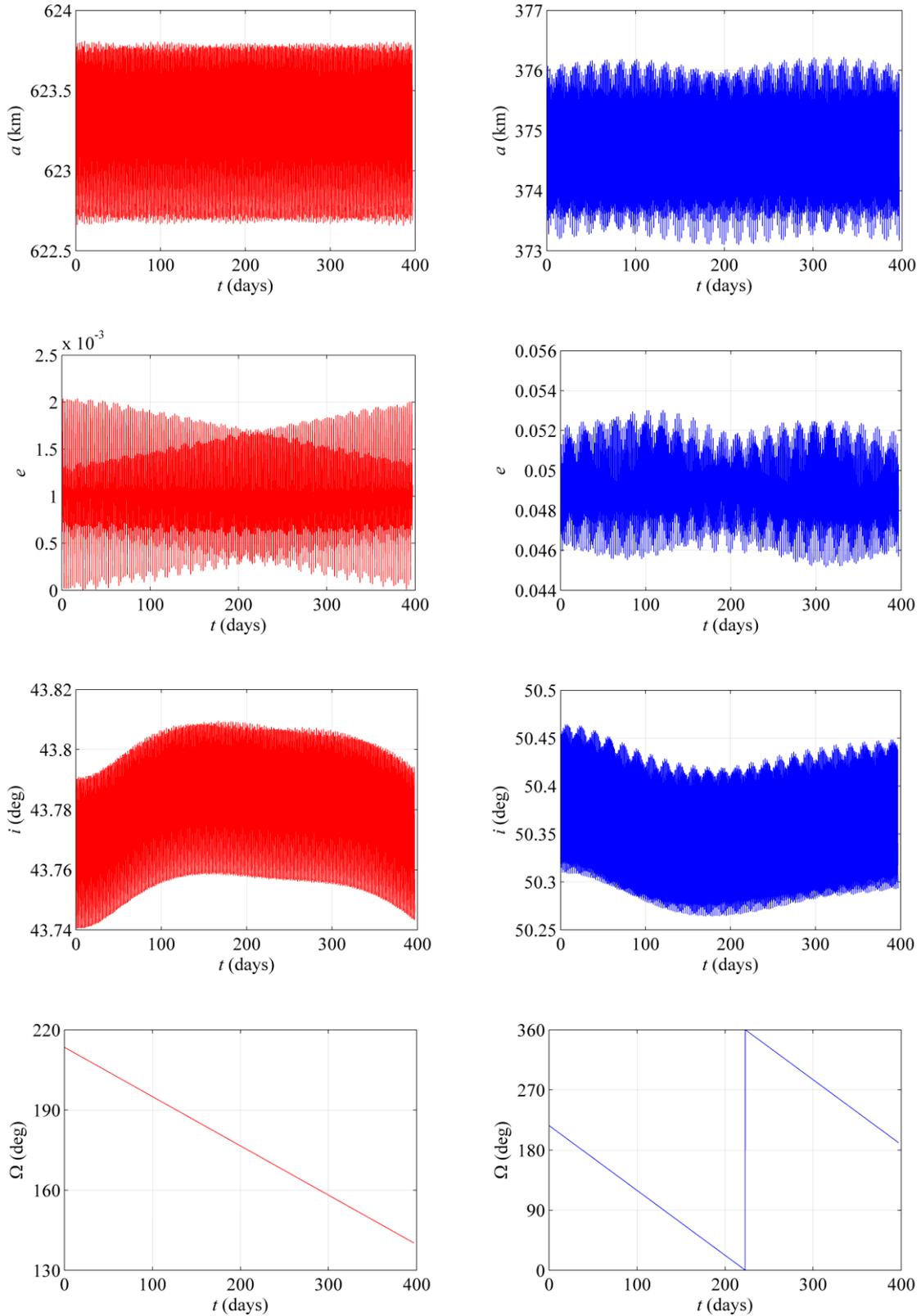

Fig. 9 Dynamical calculation of the triple-asteroid system 93 Minerva: the dark background figure shows the orbits of the two moonlets Aegis (Minerva I) and Gorgoneion (Minerva II), and the colorbar indicates the speed; the two columns show the semi-major axis, eccentricity, inclination, and the longitude of the ascending node for Aegis (left) and Gorgoneion (right).

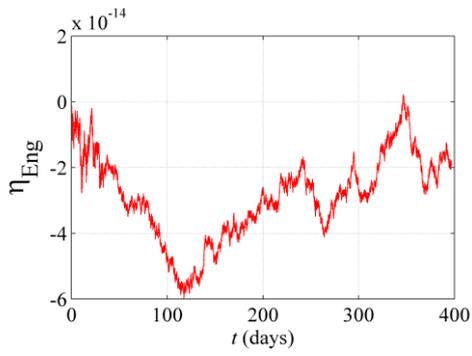 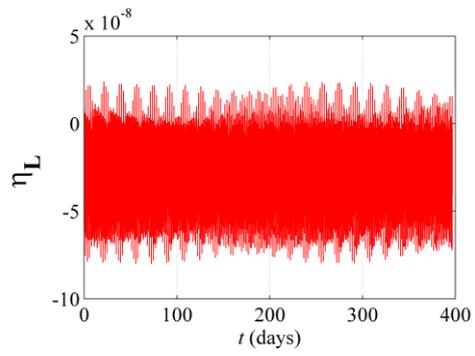

(a)                                          (b)

Fig. 10 The error of the energy and the error of the angular momentum for the calculation of the triple-asteroid system 93 Minerva. (a) shows the relative energy while (b) shows the relative angular momentum.

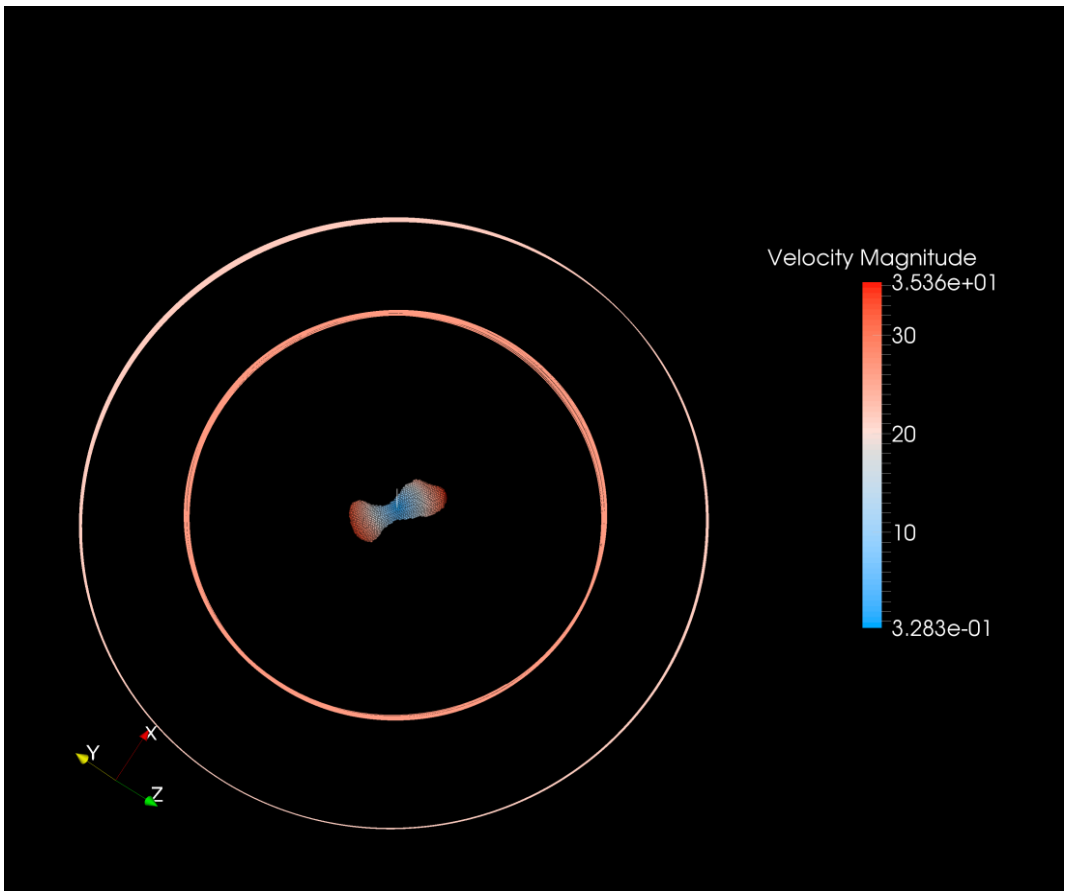

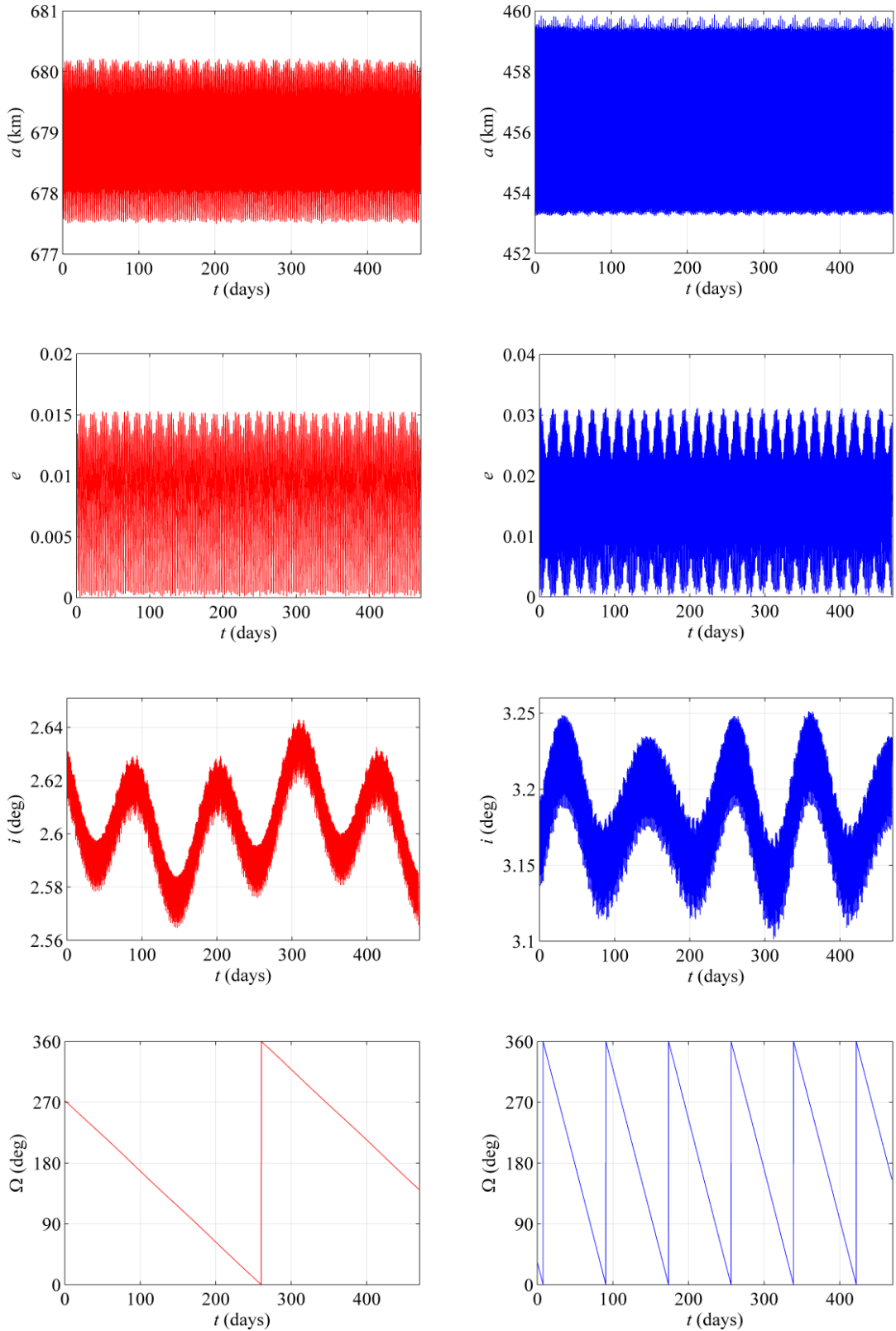

Fig. 11 Dynamical calculation of the triple-asteroid system 216 Kleopatra: the dark background figure shows the orbits of the two moonlets Alexhelios and Cleoselene, and the colorbar indicates the speed; the two columns show the semi-major axis, eccentricity, inclination, and the longitude of the ascending node for Alexhelios (left) and Cleoselene (right).

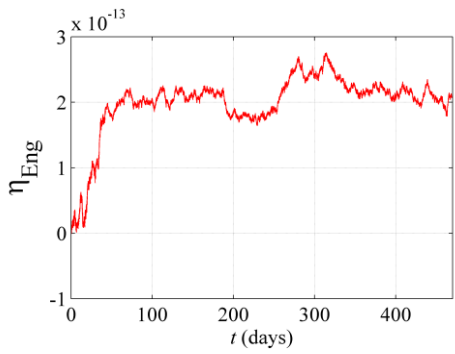 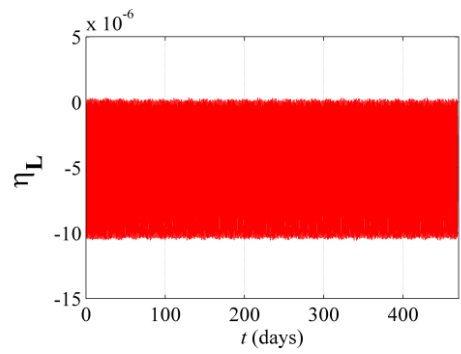

(a) (b)

Fig. 12 The error of the energy and the error of the angular momentum for the calculation of the triple-asteroid system 216 Kleopatra. (a) shows the relative energy while (b) shows the relative angular momentum.

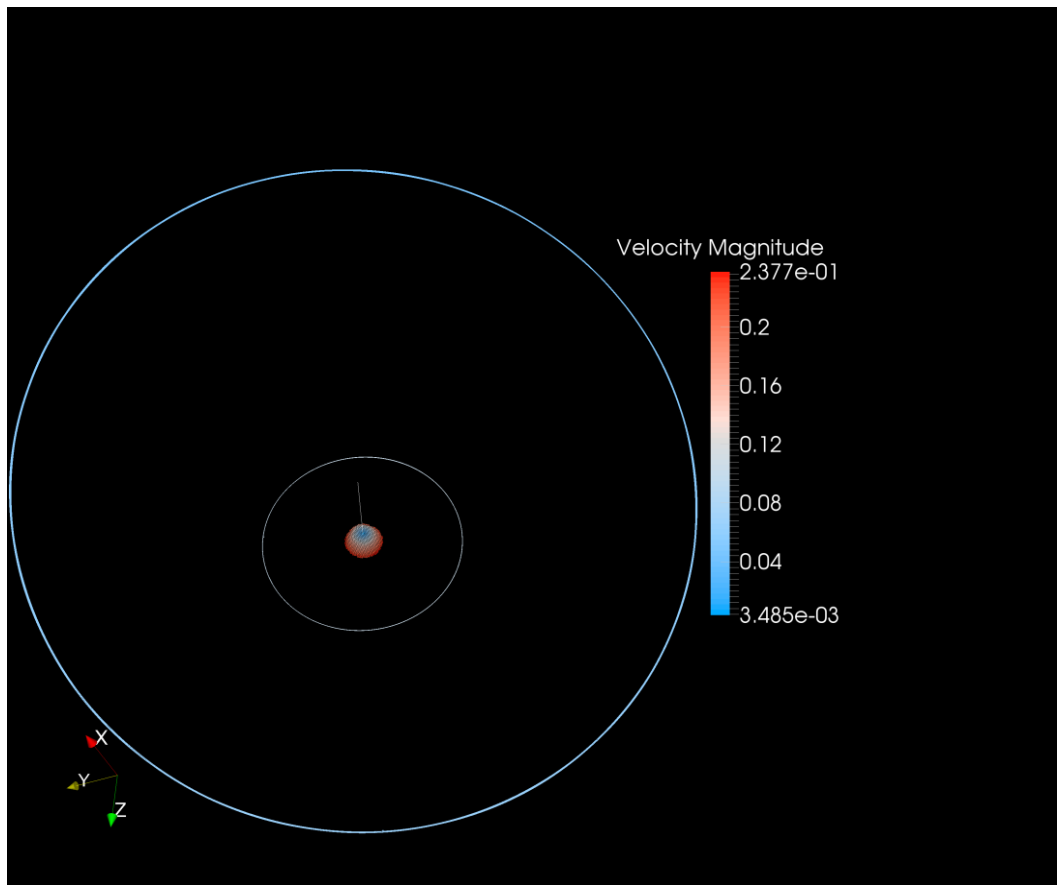

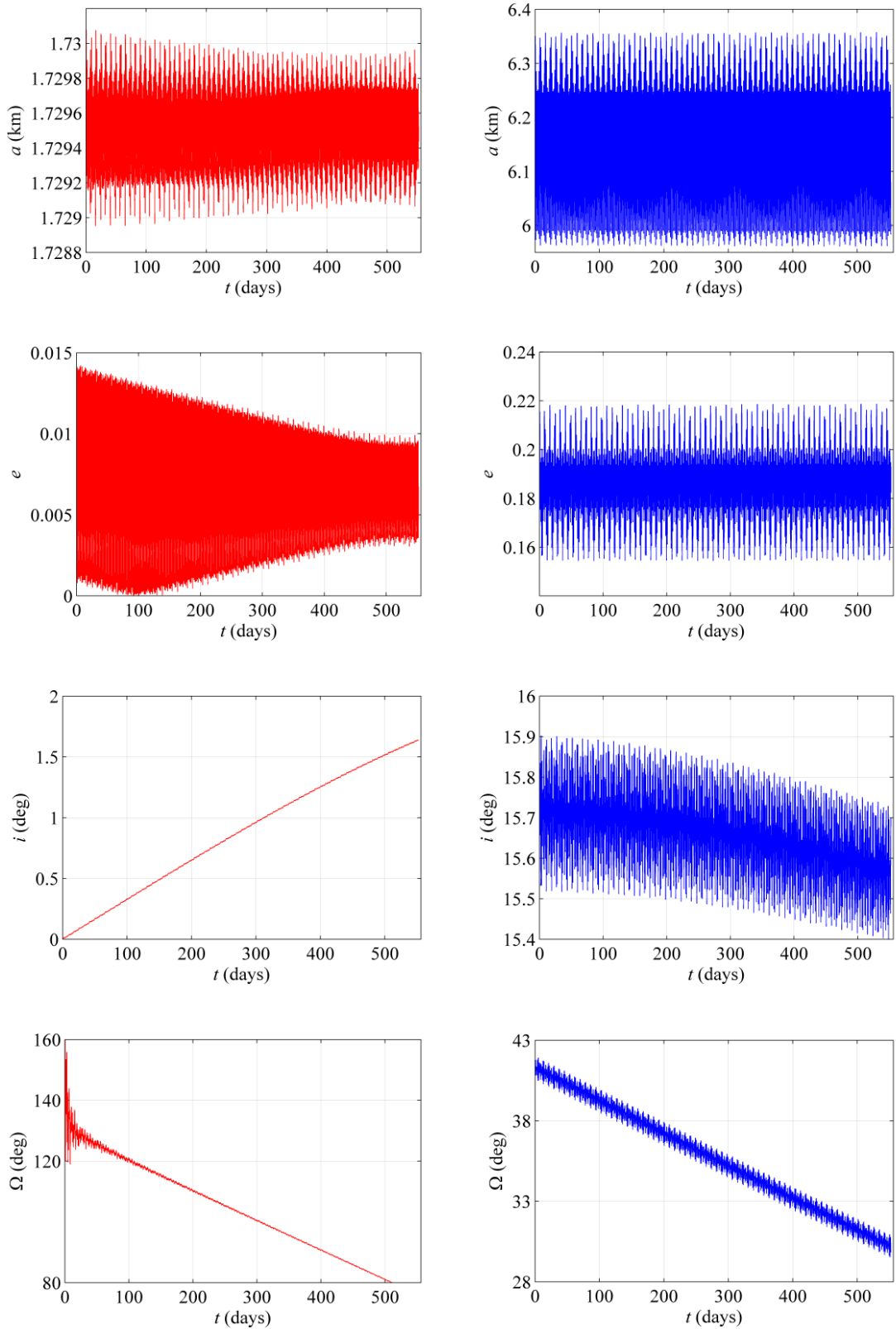

Fig. 13 Dynamical calculation of the triple-asteroid system 136617 1994CC: the dark background figure shows the orbits of the two moonlets Beta and Gamma, and the colorbar indicates the speed; the two columns show the semi-major axis, eccentricity, inclination, and the longitude of the ascending node for Beta (left) and Gamma (right).

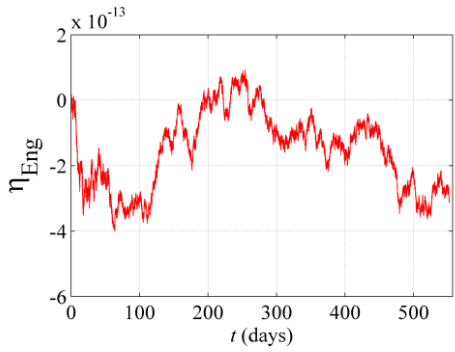 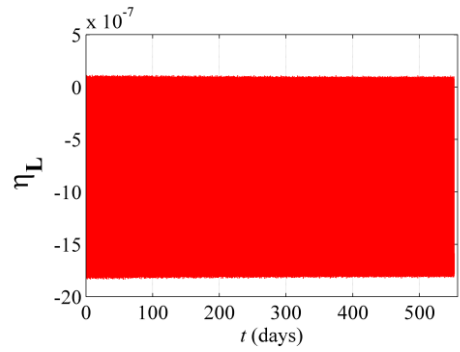

(a)                                 (b)

Fig. 14 The error of the energy and the error of the angular momentum for the calculation of the triple-asteroid system 136617 1994CC. (a) shows the relative energy while (b) shows the relative angular momentum.

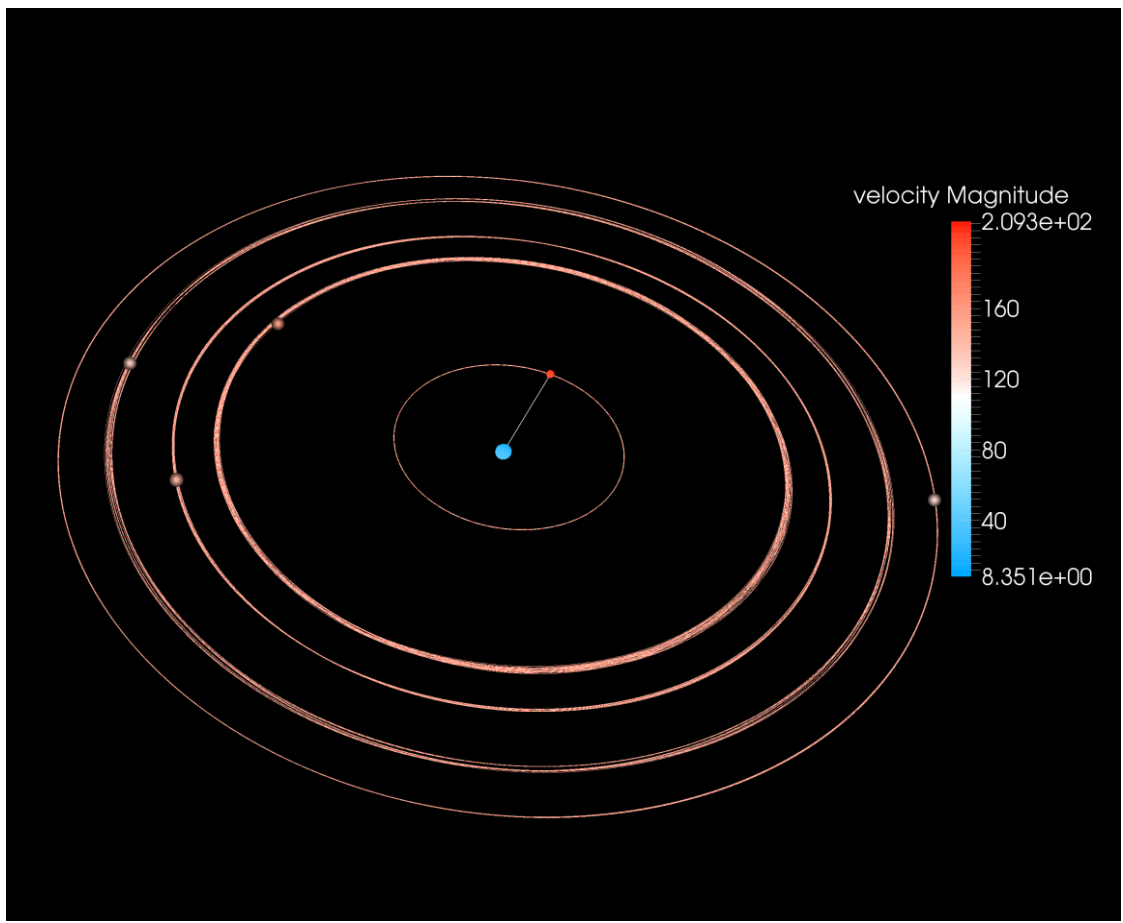

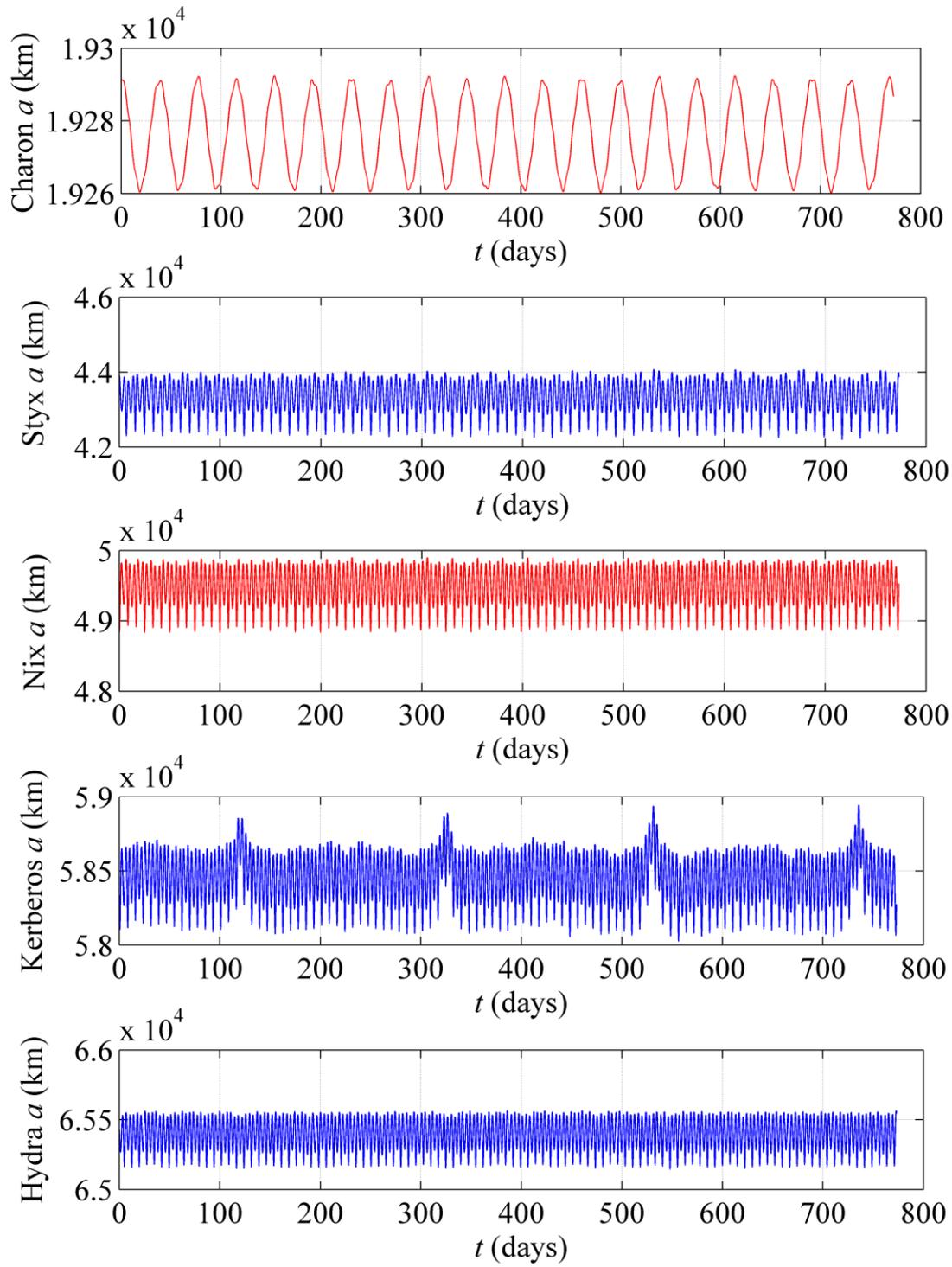

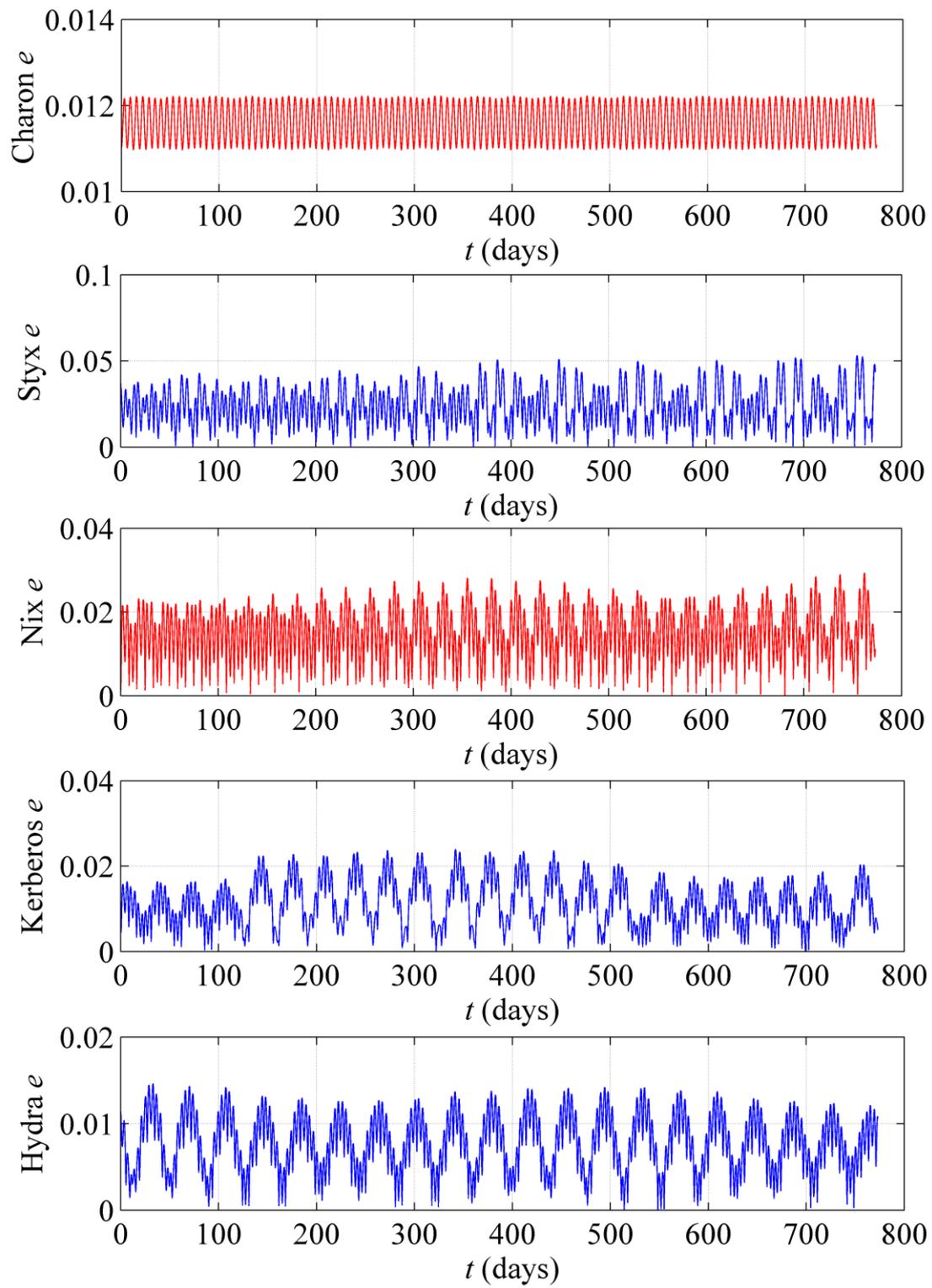

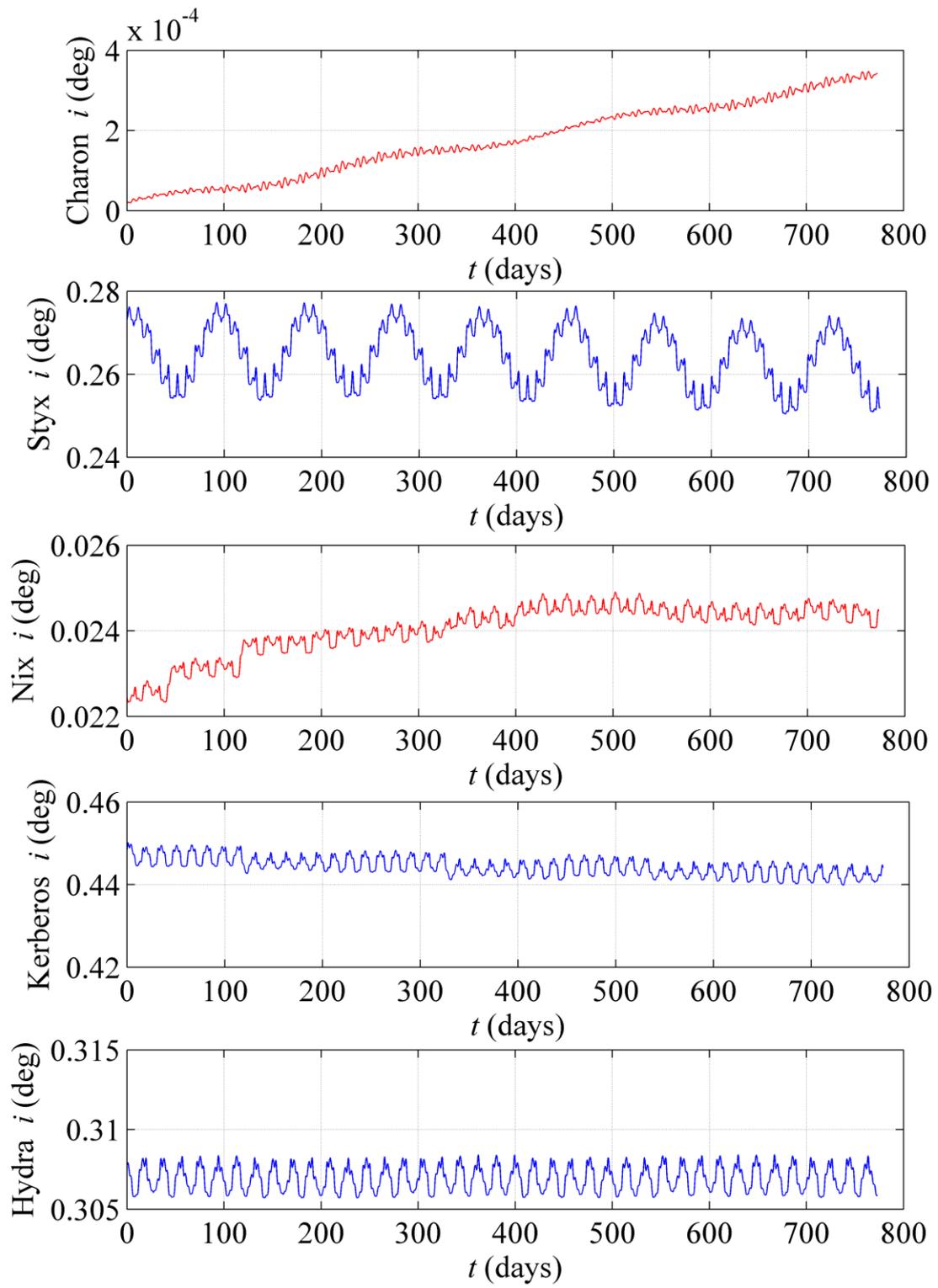

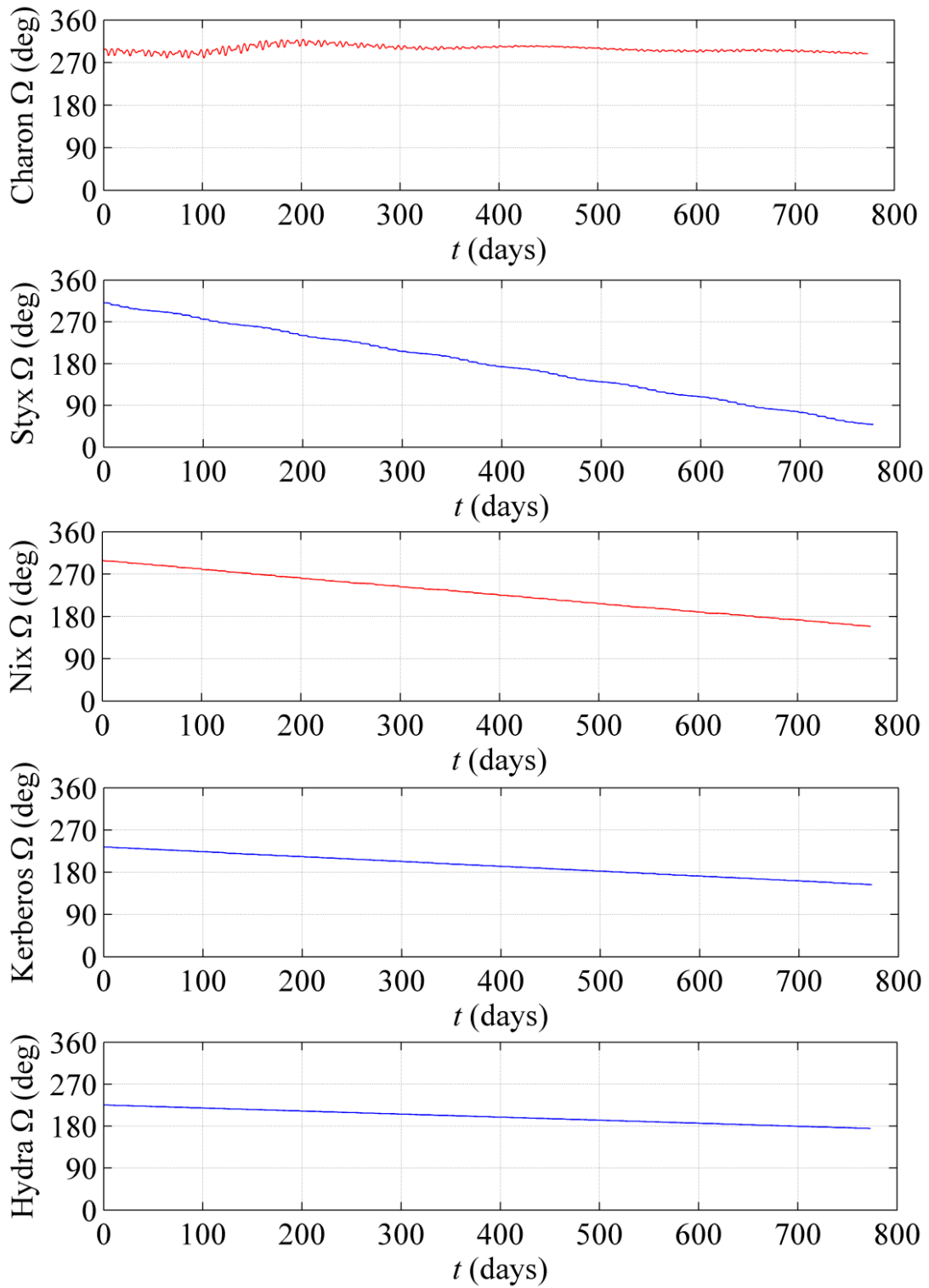

Fig. 15 Dynamical calculation of the six-body system 134340 Pluto: the semi-major axis, eccentricity, inclination, and the longitude of the ascending node for Charon, Styx, Nix, Kerberos, and Hydra are presented. Charon's orbit parameters are expressed in terms of the centroid coordinates of Pluto, while the orbit parameters of the other four moonlets are expressed in terms of the centroid inertial frame of the six-body system.

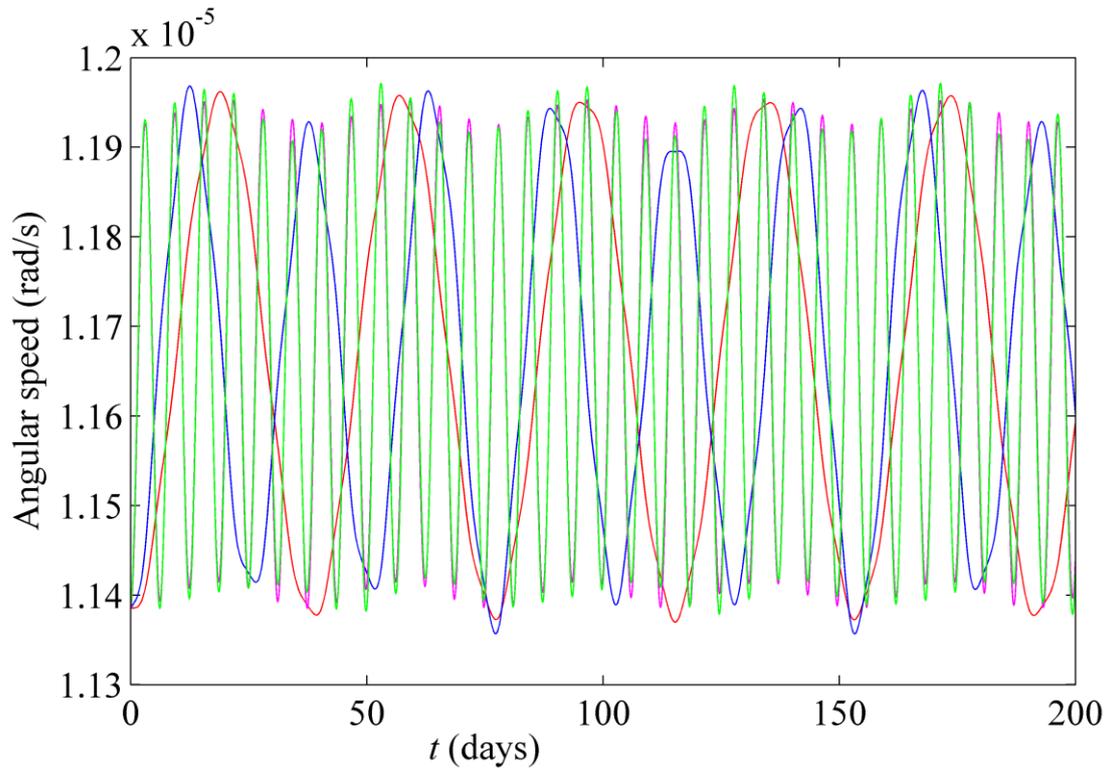

Fig. 16 Angular speeds of Pluto and Charon: the red line represents Pluto's attitude angular speed, the blue line represents Charon's attitude angular speed, the green line represents Pluto's orbital angular speed, and the magenta line represents Charon's orbital angular speed.

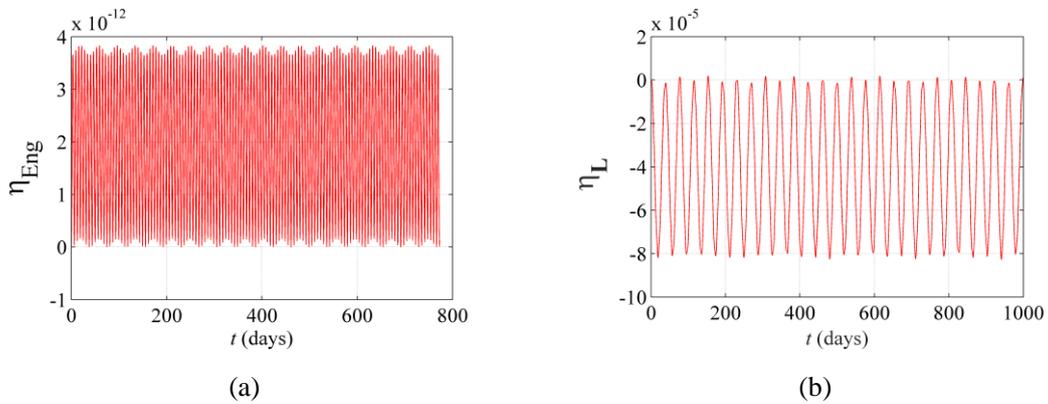

(a)  (b)

Fig. 17 The error of the energy and the error of the angular momentum for the calculation of six-body system 134340 Pluto. (a) shows the relative energy while (b) shows the relative angular momentum.

Figs. 5 and 6 shows the dynamical calculation of 45 Eugenia. The variational range of $a$ for the moonlet denoted as Princesse is 1164.05 km to 1164.7 km, giving an amplitude of about 0.65 km. The variational range of $a$ for the moonlet denoted as Petit-Prince is 610.2 km to 612.6 km, giving an amplitude of about 2.4 km. The frequency of the eccentricity of Princesse is less than that of Petit-Prince, which is determined by the orbital period. The inclinations of Princesse and Petit-Prince are periodic, and are in the intervals [9.125°, 9.18°] and [18.145°, 18.302°], respectively. The rate of change of $\Omega$ for Princesse is smaller than that for Petit-Prince. This result follows the conclusions forecasted in Section 3.5. The error of the energy is computed by the relative difference of the energy scaled by the initial value. The error of the angular momentum is computed by the relative difference of the angular momentum scaled by the initial value. Marchis et al. (2010) also integrated the orbits of the triple-asteroid system 45 Eugenia. The orbital elements of two moonlets are expressed relative to the primary's equator, which are the same as ours. Here the orbital elements of two moonlets are approximately the same as that of Marchis et al. (2010). The value looks a little different because our gravitational model for Eugenia is the polyhedron model filled by several spheres. The polyhedron model for Eugenia has 1022 vertices and 2040 faces. The discrete element model for Eugenia has 6614 spheres with the radius 3967.2m. Our model is combined by the polyhedron model and the discrete element model. The gravitational model for Eugenia that Marchis et al. (2010) used is the mass point and the J2 perturbation. Besides, most of the multiple-asteroid systems have large voids in the primary's body (Marchis et al 2010; Descamps et al. 2011). The mixed model combined by the polyhedron model and the discrete element model can not only represent the irregular shape of the primary, but also represent the large voids of it.

The dynamical calculation of 87 Sylvia is presented in Fig. 7 and 8. The variational range of *a* for the moonlet Romulus is 1351.32 km to 1352.25 km, giving an amplitude of about 0.93 km. The variational range of *a* for the moonlet Remus is 682.7 km to 685.2 km, giving an amplitude of about 2.5 km. The frequency of the eccentricity for Romulus is less than that of Remus. The inclinations of Romulus and Remus are very small, and their intervals are [0.2982°, 0.3008°] and [0.3533°, 0.3601°], respectively. The orbital plane of these two moonlets and the equator of the primary body of 87 Sylvia are nearly coincident. About the integration of orbital elements for two moonlets of 87 Sylvia, Fang et al. (2012) presented the integration of semi-major axis and eccentricity. Our results are different from that of Fang et al. (2012), because the initial parameters we used are newer data from Berthier et al. (2014). Berthier et al. (2014) only presented the orbital parameters of 87 Sylvia and didn't integrate the orbital elements. Fang et al. (2012) used the mass point and the J2 perturbation to calculate the gravitation of Sylvia. Our model for Sylvia is combined by the polyhedron model and the discrete element model. The other two moonlets we modeled to be mass points. The polyhedron model for Sylvia also has 1022 vertices and 2040 faces. The discrete element model for Sylvia has 7957 spheres with the radius 5699m.

The dynamical calculation of 93 Minerva is presented in Figs. 9 and 10. The variational range of *a* for the moonlet Aegis is 622.65 km to 623.8 km, giving an amplitude of about 1.15 km. The variational range of *a* for the moonlet Gorgoneion is 373.25 km to 376.1 km, giving an amplitude of about 2.85 km. The inclinations of the moonlets of 93 Minerva are much greater than those of the other triple-asteroid systems considered. The inclinations of Aegis and Gorgoneion are in the intervals [43.74°, 43.81°] and [50.27°, 50.465°], respectively. Marchis et al. (2013) calculated the orbital elements for two moonlets of 93 Minerva respect to EQJ2000, the inclinations of Aegis and Gorgoneion are expressed relative to Earth's equator. They didn't present the integration of orbital elements for Aegis and Gorgoneion relative to Minerva. Here we used the initial data from Marchis et al. (2013) and integrated the orbital elements for Aegis and Gorgoneion relative to Minerva's equator. Our model for Minerva is combined by the polyhedron model and the discrete element model. The other two moonlets we modeled to be mass points. The polyhedron model for Minerva has 402 vertices and 800 faces. The discrete element model for Minerva has 12582 spheres with the radius 2901m.

The dynamical calculation of 216 Kleopatra is presented in Figs. 11 and 12. The variational range of $a$ for the moonlet Alexhelios is 677.5 km to 680.2 km, giving an amplitude of about 2.7 km. The variational range of $a$ for the moonlet Cleoselene is 453.2 km to 459.8 km, giving an amplitude of about 6.6 km. The inclinations of Alexhelios and Cleoselene are periodic and in the intervals [2.565°, 2.643°] and [3.11°, 3.25°], respectively. Descamps et al. (2011) calculated the orbital elements for two moonlets of 216 Kleopatra respect to ECJ2000, the inclinations of Alexhelios and Cleoselene are expressed relative to Earth's equator. They didn't present the integration of orbital elements for Alexhelios and Cleoselene relative to Kleopatra. Here we used the initial data from Descamps et al. (2011) and integrated the orbital elements for Alexhelios and Cleoselene relative to Kleopatra's equator. Our model for Kleopatra is combined by the polyhedron model and the discrete element model. The other two moonlets we modeled to be mass points. The polyhedron model for Kleopatra has 2048 vertices and 4092 faces. The discrete element model for Kleopatra has 10339 spheres with the radius 2190m.

The dynamical calculation of 136617 1994CC is presented in Figs. 13 and 14. The variational range of *a* for the moonlet Beta is 1.72895 km to 1.7301 km, giving an amplitude of about 0.00115 km. The variational range of *a* for the moonlet Gama is 5.97 km to 6.36 km, giving an amplitude of about 0.39 km. Relative to the other triple-asteroid systems considered, 136617 1994CC is not a high size ratio system. Figure 13 indicates that, when the inclination of Beta is around zero, the value of $\Omega$ for Beta exhibits a large amplitude. Fang et al. (2011) presented the orbital elements for two moonlets respect to equatorial frame of J2000, the inclinations of Beta and Gama are expressed relative to Earth's equator. They didn't present the integration of orbital elements for Beta and Gama relative to Alfa. Here we used the initial data of orbits from Fang et al. (2011) as well as mass and shape data from Brozović et al. (2011), and integrated the orbital elements for Beta and Gama relative to Alfa's equator. Our model for Alfa is the ellipsoid model, and for Beta and Gama are mass point model.

Figs. 15-17 show the dynamical calculation of 134340 Pluto. The results indicate that Pluto and Charon are gravitationally locked, and some drifting is evident for the attitude of these two bodies. Fig. 16 shows the angular speeds of the orbits and attitudes of Pluto and Charon. From Fig. 16, we can conclude that the attitude angular speeds of Pluto and Charon are not constant, and both exhibit a periodic variation. Showalter and Hamilton (2015) integrated the orbits of the six-body system 134340 Pluto; they investigated the Styx-Nix-Hydra resonance and the rotation of Nix, the results show that the Pluto and Charon make the moons rotate chaotically. Here we calculated the orbit motion and attitude of Pluto and Charon, the results show that the four parameters, the orbital angular speed and attitude angular speed of Pluto and Charon are not constant. The orbital angular speeds of Pluto and Charon have identical periods and are nearly coincident, while the period of the attitude angular speed of Pluto is greater than that of Charon. In addition, the other four moonlets, Styx, Nix, Kerberos, and Hydra, produce perturbations in the attitudes of Pluto and Charon, which cause the observed variations in their attitudes. The angular momentum of the system is $6.85 \times 10^{10}$ J·s.

From the eccentricity of Charon, which is in the interval of [0.01096°, 0.01223°], we know that the orbit of Charon relative to Pluto is not strictly circular. The values of *a* for the other four moonlets are observed to increase in turn while the eccentricities decrease, and the intervals of the eccentricities for Styx, Nix, Kerberos, and Hydra are [0.0, 0.05], [0.0, 0.03], [0.0, 0.025], and [0.0, 0.015], respectively. The period of Styx's inclination is 91.25 days and the period of Hydra's inclination is 19.2 days, and the inclination of Kerberos is greater than those of the other moonlets. The orbital planes of Styx, Nix, Kerberos, and Hydra are nearly coincident with the equator of the gravitationally-locked Pluto-Charon system.

For the moonlets of the high size ratio triple-asteroid systems 45 Eugenia, 87 Sylvia, 93 Minerva, and 216 Kleopatra and the four moonlets of the Pluto-Charon system, the rate of change of $\Omega$ for the inner moonlet is greater than that for the outer moonlet.

All the moonlets have small inclinations except for those of 93 Minerva, where, as discussed, the inclinations of Aegis and Gorgoneion are intervals of $[43.74^o, 43.8^o]$ and $[50.27^o, 50.46^o]$, respectively. The two moonlets of the high size ratio triple-asteroid systems exhibit changing inclinations, which verifies the conclusion given in Section 3.5. In addition, the two moonlets of 87 Sylvia and the four moonlets of the gravitationally-locked Pluto-Charon system have the smallest inclinations of all the systems considered, where all are less than $0.5^o$. The inclinations of the moonlets Alexhelios and Cleoselene of 216 Kleopatra change by $0.078^o$ and $0.14^o$, respectively, and the inclinations of the moonlets Princesse and Petit-Prince of 45 Eugenia change by about $0.055^o$ and $0.157^o$, respectively. For the four high size ratio triple-asteroid systems 45 Eugenia, 87 Sylvia, 93 Minerva, and 216 Kleopatra, the outer moonlets have smaller inclinations than the inner moonlets. However, the size ratio of 136617 1994CC is smaller, and the inclinations of its moonlets exhibit a different tendency, where the outer moonlet has a larger inclination than that of the inner one.

## 4.4 Relation Diagram of Orbital Elements

In this section, we investigate the relation diagram of orbital elements for these five triple-asteroid systems and the six-body system 134340 Pluto. Figs. 18 to 22 show the relation diagram of moonlets' orbital elements for 45 Eugenia, 87 Sylvia, 93 Minerva, 216 Kleopatra, and 136617 1994CC. Fig. 23 shows the relation diagram of moonlets' orbital elements for 134340 Pluto.

From Fig. 18, one can see that the region of the relation diagram of $a$ and $e$ for 45 Eugenia's moonlets looks like a quadrangle and has a serrated edge. From Fig. 19, one can see that the region of the relation diagram of $a$ and $e$ for 87 Sylvia's moonlets looks like a trapezium. While for 216 Kleopatra's moonlets, the region looks like a cuboid. From Figs. 20 and 22, one can see that the region of the relation diagram of $a$ and $e$ for moonlets of 93 Minerva and 136617 1994CC are quite different from that for 45 Eugenia, 87 Sylvia, and 216 Kleopatra. The region of the relation diagram of $a$ and $e$ for moonlets 1 of 93 Minerva has a small hollow region. From Fig. 23, one can see that the region of the relation diagram of $a$ and $e$ for Charon is different than for that of other four moonlets. The regions of the relation diagram of $a$ and $e$ for the moonlets of these five triple-asteroid systems and the six-body system 134340 Pluto don't have any straggling points, which means that the motion of $a$ and $e$ for the moonlets are stable. Previous works have investigated the stability regions around the moonlets of triple-asteroid 2001 SN263 (Araujo et al. 2012) and instability regions around the moonlets of triple-asteroid 87 Sylvia (Frouard and Compère 2012). These studies plotted the relation diagram of orbital elements for a massless particle in the triple-asteroid systems, the primary is assumed to be a mass point with the J2 perturbation, and the two moonlets are assumed to be mass points. The irregular shape of the primary is neglected. Here we investigate the relation diagram of orbital elements for moonlets, not a massless particle. Jiang et al. (2015c) found a stable region consiste of retrograde periodic orbits; the periodic orbits are also nearly circular and have zero inclination. The result presented a point of view of the motion stability for Alexhelios and Cleoselene of the large size ratio triple-asteroid system 216 Kleopatra. However, Jiang et al. (2015c) only considered the polyhedral model of the primary and neglected the gravitation of the two moonlets. Here we considered the gravitation caused by the irregular shape of all the primaries of the triple-asteroid systems except 136617 1994CC, the moonlets' gravitation are also calculated.

Among these five triple-asteroid systems, the figures of the relation diagram of $a$ and $e$ for 93 Minerva's two moonlets are quite different. The figure of the relation diagram of $a$ and $e$ for moonlet 1 of 93 Minerva has a hollow region, while the figure of the relation diagram of $a$ and $e$ for moonlet 2 of 93 Minerva has a solid region. The figures of the relation diagram of $a$ and $e$ for 136617 1994CC's two moonlets are also quite different. The figure of the relation diagram of $a$ and $e$ for moonlet 1 of 136617 1994CC looks more dense in central region than boundary region. However, the boundary region of the figure for moonlet 2 of 136617 1994CC looks also dense.

From the figures of the relation diagram of $i$ and $\Omega$ for moonlets of these five triple-asteroid systems, one can see that the value of $\Omega$ for moonlets of 45 Eugenia, 87 Sylvia, 93 Minerva, and 216 Kleopatra can vary from $0^o$ to $360^o$. The figures of the relation diagram of $i$ and $\Omega$ for moonlets of 45 Eugenia, 87 Sylvia, 93 Minerva, and 216 Kleopatra seem to be periodic. For 136617 1994CC, the figures of the relation diagram of $i$ and $\Omega$ for moonlets are quite different from that of other four triple-asteroid systems. For some moonlets of theses triple-asteroid systems we calculated, the variety period of Ω is greater than the compute time. For example, the compute time for the dynamical configuration of 93 Minerva is 400d, but the variety period of Ω for the moonlet 1 of 93 Minerva is greater than 400d. About the relation diagram of $i$ and Ω for moonets of 134340 Pluto, the figures of the relation diagram of $i$ and Ω for Charon and Styx are special. In the figure of the relation diagram of $i$ and Ω for Charon, the local trajectory looks like a circle but not a closed circle. In the figure of the relation diagram of $i$ and Ω for Styx, the local trajectory has two short-periodic terms.

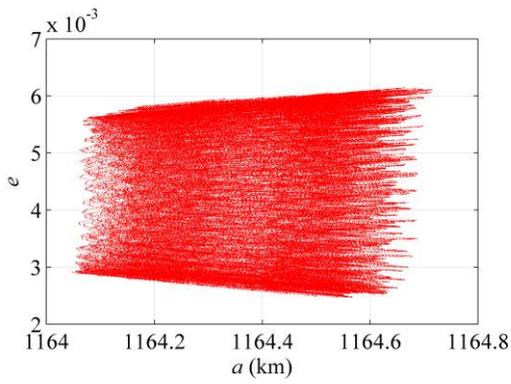

(a)

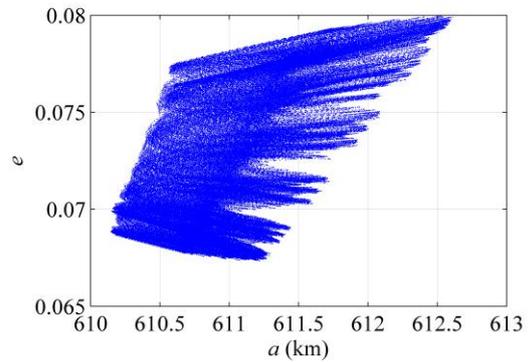

(b)

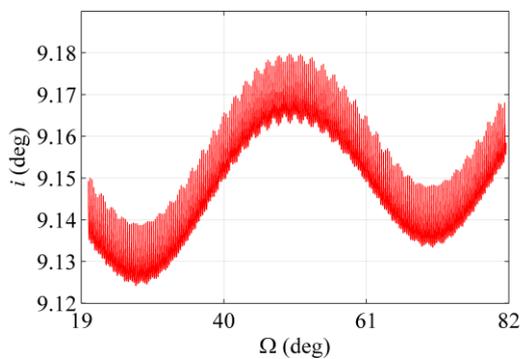

(c)

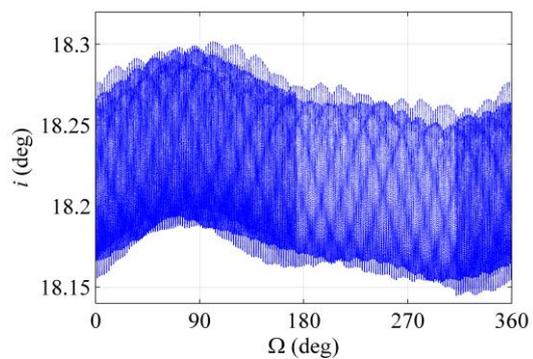

(d)

Fig. 18 Relation diagram of orbital elements for the triple-asteroid system 45 Eugenia

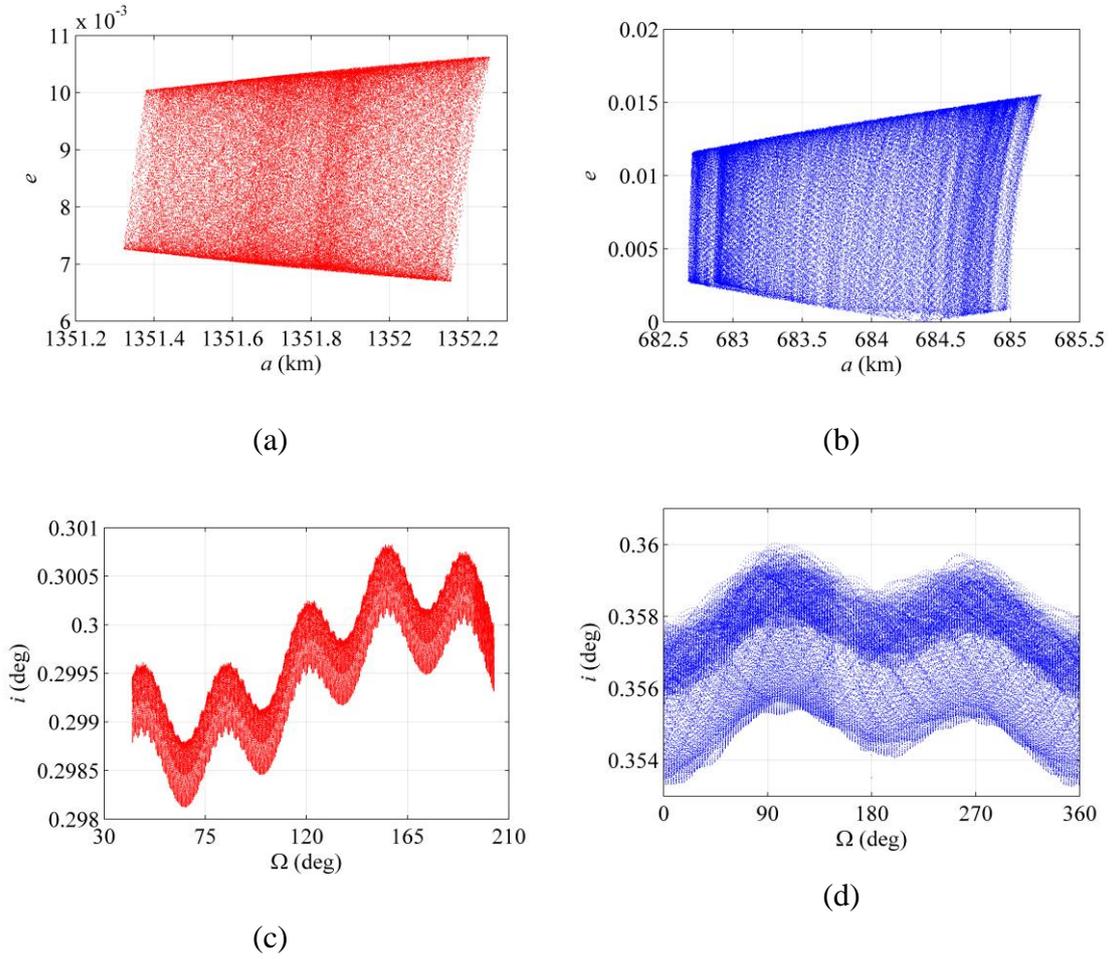

(a)

(b)

(c)

(d)

Fig. 19 Relation diagram of orbital elements for the triple-asteroid system 87Sylvia

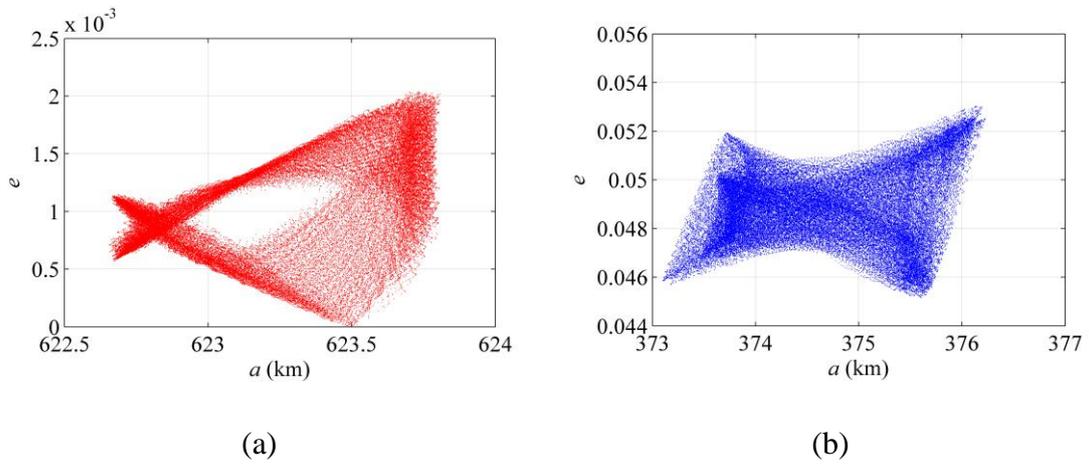

(a)

(b)

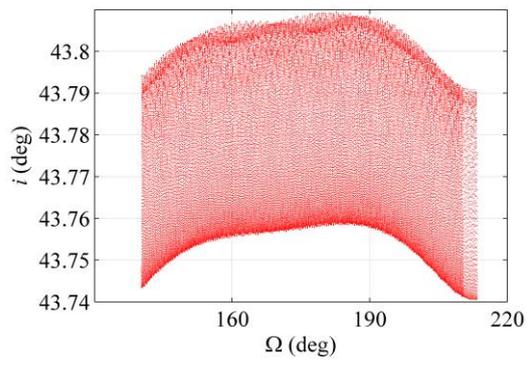
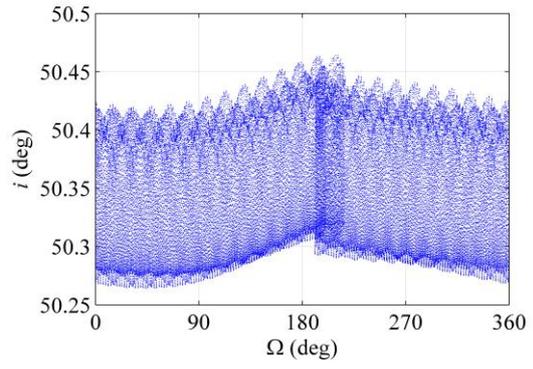

(c)             (d)

Fig. 20 Relation diagram of orbital elements for the triple-asteroid system 93Minerva

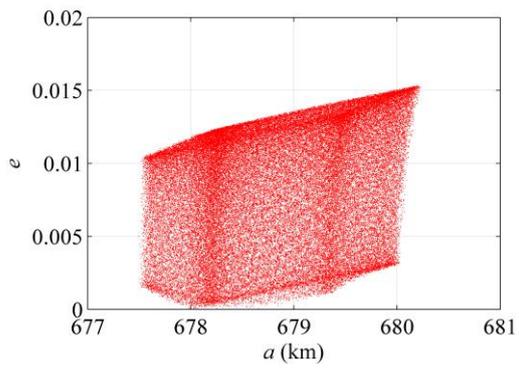
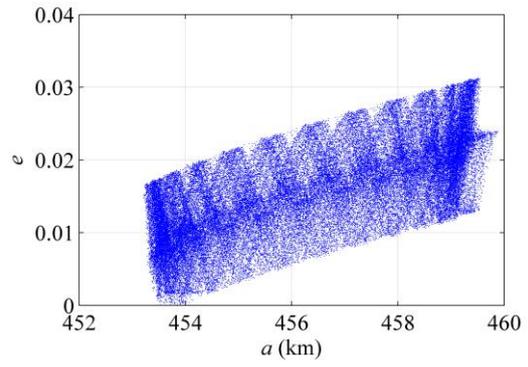

(a)             (b)

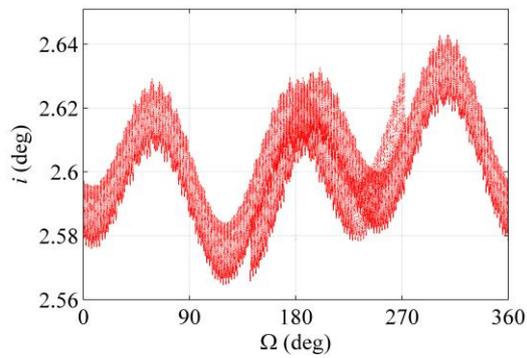
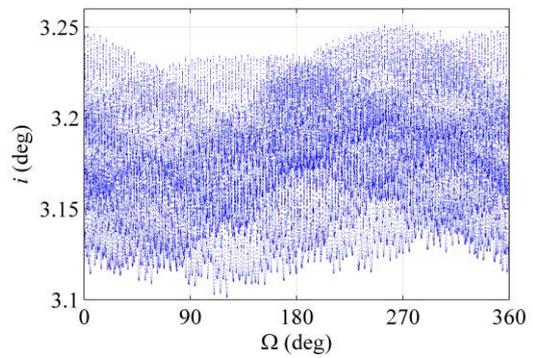

(c)             (d)

Fig. 21 Relation diagram of orbital elements for the triple-asteroid system 216Kleopatra

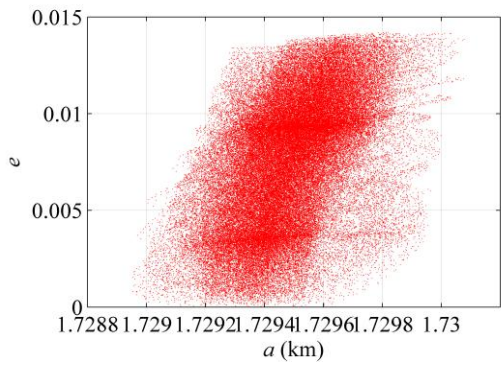
(a)

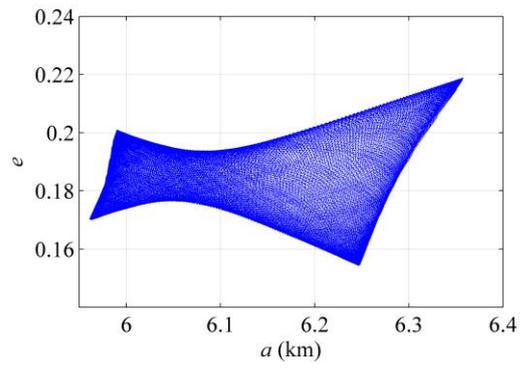
(b)

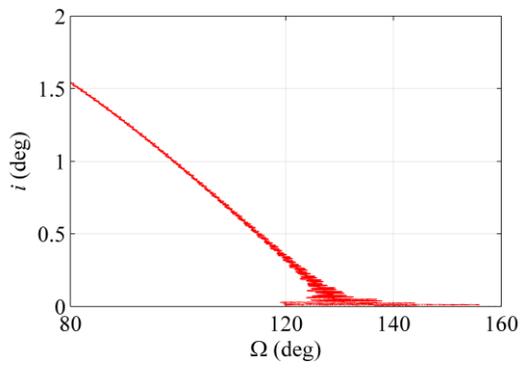
(c)

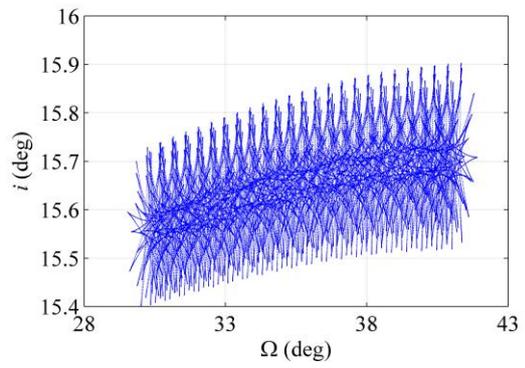
(d)

Fig. 22 Relation diagram of orbital elements for the triple-asteroid system 136617 1994CC

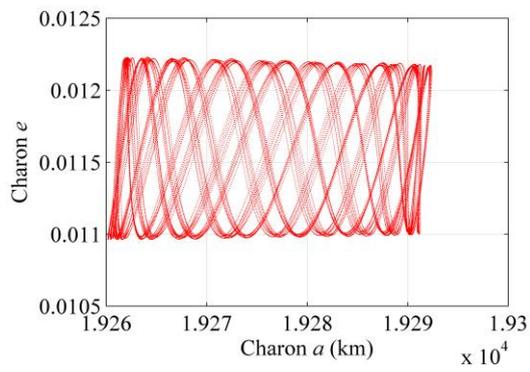

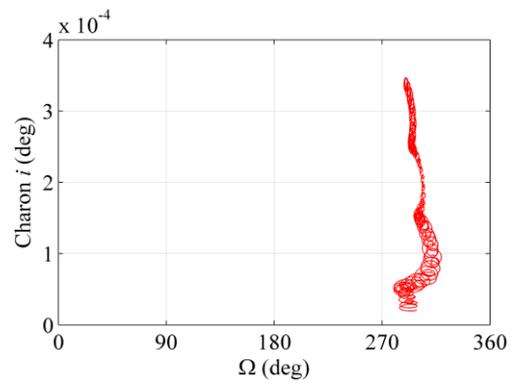

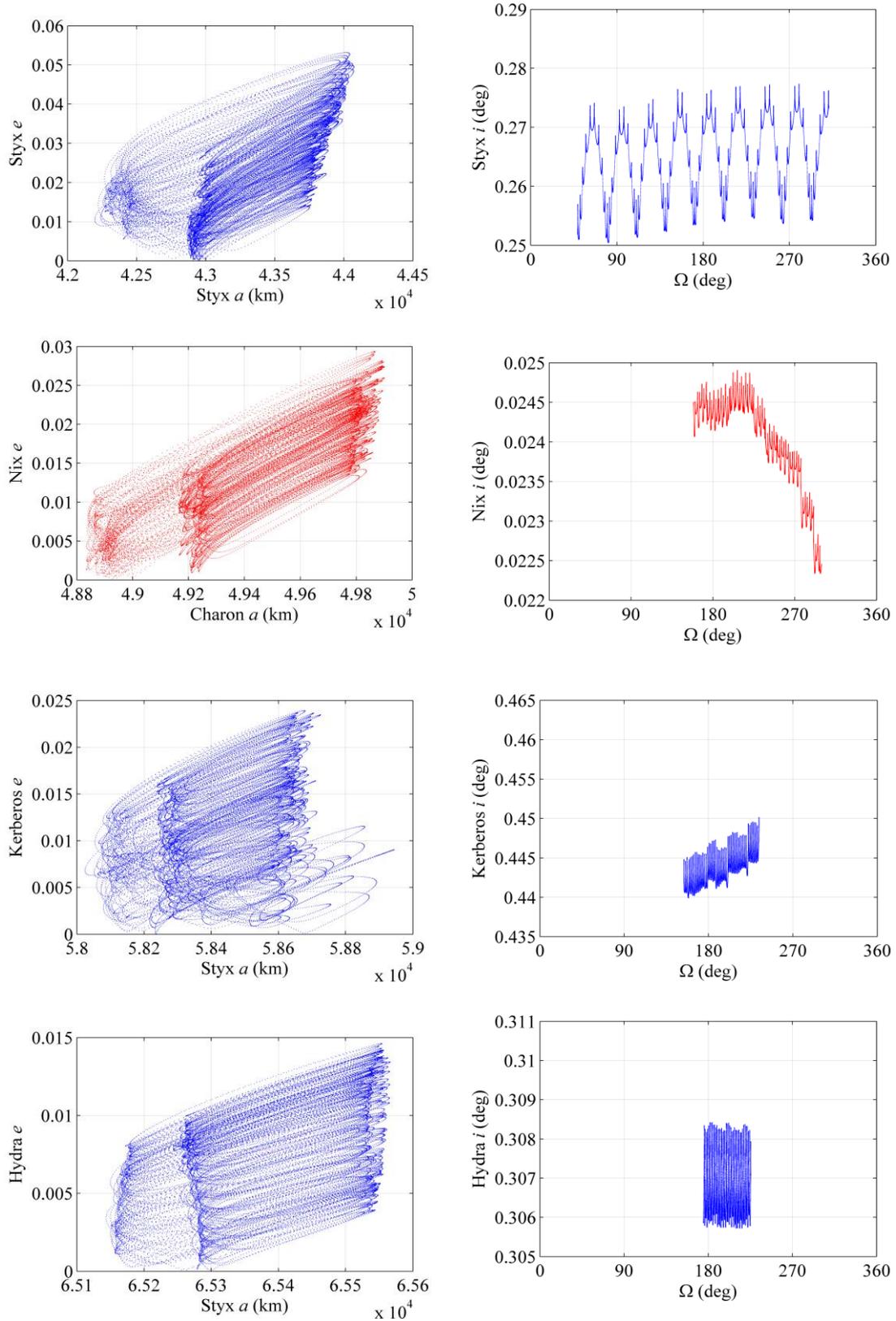

Fig. 23 Relation diagram of orbital elements for the six-body system 134340 Pluto

## 5. Conclusions

In this paper, we derived the mutual gravitational potential, static electric potential, and magnetic potential of a celestial system comprised of $n$ irregular bodies. The forces and torques of gravity, static electric, and magnetism acting on each body of the system were presented. Using the forces and torques, dynamical equations of the system were established, and expressed in terms of the inertial space and the body-fixed frame. The Jacobi integral and two other conservative values of the system were presented. The equation for the equilibrium of the system was derived from the dynamical equations relative to the body-fixed frame. The equilibrium conditions and the stability of the relative equilibrium were analyzed.

    The dynamical configurations of five triple-asteroid systems 45 Eugenia, 87 Sylvia, 93 Minerva, 216 Kleopatra, and 136617 1994CC and the six-body system 134340 Pluto were calculated. Moonlets Aegis and Gorgoneion of 93 Minerva exhibited the largest inclinations, which are 43.77 ° and 50.35 °, respectively. The orbital angular speed and attitude angular speed of the gravitationally-locked bodies Pluto and Charon of 134340 Pluto were shown to vary periodically. For 45 Eugenia's moonlets, the figure of the relation diagram of $a$ and $e$ looks like a quadrangle and has a serrated edge. For 87 Sylvia's moonlets, the figure looks like a trapezium. For 216 Kleopatra's moonlets, it looks like a cuboid. For 93 Minerva's moonlet 1, the figure of the relation diagram of $a$ and $e$ has a hollow region. For the six-body system 134340 Pluto, the relation diagram of orbital elements of different moonlets are different.

**Acknowledgements**

This research was supported by the National Natural Science Foundation of China (No. 11372150), the National Basic Research Program of China (973 Program, 2012CB720000), and the State Key Laboratory of Astronautic Dynamics Foundation (No. 2015ADL-DW02).

**Appendix A. The static electric potential and magnetic potential**

**A1. The potential and dynamic equation relative to the inertial space**

In this section, we present the static electric potential and magnetic potential of multiple irregular bodies, as well as the dynamic equation relative to the inertial space.

The total static electric potential energy can be written as

$$U_{se} = -\sum_{k=1}^{n-1}\sum_{j=k+1}^{n} \frac{1}{4\pi\varepsilon_0} \int_{\beta_k}\int_{\beta_j} \frac{\rho_{se}(\mathbf{D}_k)\rho_{se}(\mathbf{D}_j)dV(\mathbf{D}_j)dV(\mathbf{D}_k)}{\|\mathbf{A}_k\mathbf{D}_k - \mathbf{A}_j\mathbf{D}_j + \mathbf{r}_k - \mathbf{r}_j\|}, \tag{A1}$$

and the total magnetic potential energy can be written as

$$U_m = -\sum_{k=1}^{n-1}\sum_{j=k+1}^{n} \frac{\mu_0}{4\pi} \int_{\beta_k}\int_{\beta_j} \frac{\mathbf{J}(\mathbf{D}_k)\cdot\mathbf{J}(\mathbf{D}_j)dV(\mathbf{D}_j)dV(\mathbf{D}_k)}{\|\mathbf{A}_k\mathbf{D}_k - \mathbf{A}_j\mathbf{D}_j + \mathbf{r}_k - \mathbf{r}_j\|}. \tag{A2}$$

Thus, the total energy of the system becomes

$$H = T + U_g + U_e + U_m = \frac{1}{2}\sum_{k=1}^{n}\left(m_k\|\dot{\mathbf{r}}_k\|^2 + \langle\Phi_k,\mathbf{I}_k\Phi_k\rangle\right)$$
$$-\sum_{k=1}^{n-1}\sum_{j=k+1}^{n}\int_{\beta_k}\int_{\beta_j}\left[G\rho_g(\mathbf{D}_k)\rho_g(\mathbf{D}_j) + \frac{1}{4\pi\varepsilon_0}\rho_{se}(\mathbf{D}_k)\rho_{se}(\mathbf{D}_j) + \frac{\mu_0}{4\pi}\mathbf{J}(\mathbf{D}_k)\cdot\mathbf{J}(\mathbf{D}_j)\right]\frac{dV(\mathbf{D}_j)dV(\mathbf{D}_k)}{\|\mathbf{A}_k\mathbf{D}_k - \mathbf{A}_j\mathbf{D}_j + \mathbf{r}_k - \mathbf{r}_j\|}$$

(A3)

The total static electric force acting on $\beta_k$ can be written as

$$\mathbf{f}_{se}^k = -\frac{1}{4\pi\varepsilon_0}\sum_{j=1,j\neq k}^{n}\int_{\beta_k}\int_{\beta_j} \frac{(\mathbf{A}_k\mathbf{D}_k - \mathbf{A}_j\mathbf{D}_j + \mathbf{r}_k - \mathbf{r}_j)}{\|\mathbf{A}_k\mathbf{D}_k - \mathbf{A}_j\mathbf{D}_j + \mathbf{r}_k - \mathbf{r}_j\|^3}\rho_{se}(\mathbf{D}_k)\rho_{se}(\mathbf{D}_j)dV(\mathbf{D}_j)dV(\mathbf{D}_k), \tag{A4}$$

and the total magnetic force acting on $\beta_k$ can be written as

$$\mathbf{f}_m^k = -\frac{\mu_0}{4\pi} \sum_{j=1, j\neq k}^{n} \int_{\beta_k} \int_{\beta_j} \frac{(\mathbf{A}_k \mathbf{D}_k - \mathbf{A}_j \mathbf{D}_j + \mathbf{r}_k - \mathbf{r}_j)}{\|\mathbf{A}_k \mathbf{D}_k - \mathbf{A}_j \mathbf{D}_j + \mathbf{r}_k - \mathbf{r}_j\|^3} \mathbf{J}(\mathbf{D}_k) \cdot \mathbf{J}(\mathbf{D}_j) dV(\mathbf{D}_j) dV(\mathbf{D}_k). \quad (A5)$$

The resultant static electric torque acting on $\beta_k$ can be expressed as

$$\mathbf{n}_{se}^k = \frac{1}{4\pi\varepsilon_0} \sum_{j=1, j\neq k}^{n} \int_{\beta_k} \int_{\beta_j} \frac{(\mathbf{A}_k \mathbf{D}_k + \mathbf{r}_k) \times (\mathbf{A}_j \mathbf{D}_j + \mathbf{r}_j)}{\|\mathbf{A}_k \mathbf{D}_k - \mathbf{A}_j \mathbf{D}_j + \mathbf{r}_k - \mathbf{r}_j\|^3} \rho_{se}(\mathbf{D}_k) \rho_{se}(\mathbf{D}_j) dV(\mathbf{D}_j) dV(\mathbf{D}_k),$$

(A6)

and the resultant magnetic torque acting on $\beta_k$ can be expressed as

$$\mathbf{n}_m^k = \frac{\mu_0}{4\pi} \sum_{j=1, j\neq k}^{n} \int_{\beta_k} \int_{\beta_j} \frac{(\mathbf{A}_k \mathbf{D}_k + \mathbf{r}_k) \times (\mathbf{A}_j \mathbf{D}_j + \mathbf{r}_j)}{\|\mathbf{A}_k \mathbf{D}_k - \mathbf{A}_j \mathbf{D}_j + \mathbf{r}_k - \mathbf{r}_j\|^3} \mathbf{J}(\mathbf{D}_k) \cdot \mathbf{J}(\mathbf{D}_j) dV(\mathbf{D}_j) dV(\mathbf{D}_k). \quad (A7)$$

Then the dynamical equation relative to the inertial space is now

$$\begin{cases} \dot{\mathbf{p}}_k = \mathbf{f}_g^k + \mathbf{f}_{se}^k + \mathbf{f}_m^k \\ \dot{\mathbf{r}}_k = \dfrac{\mathbf{p}_k}{m_k} \\ \dot{\mathbf{K}}_k = \mathbf{n}_g^k + \mathbf{n}_{se}^k + \mathbf{n}_m^k \\ \dot{\mathbf{A}}_k = \widehat{\boldsymbol{\psi}}_k \mathbf{A}_k \end{cases}, \quad k = 1, 2, \cdots, n, \quad (A8)$$

The three conservative values of the dynamics equation are

$$H = T + U_g + U_{se} + U_m, \quad \mathbf{p} = \sum_{k=1}^{n} \mathbf{p}_k, \quad \mathbf{K} = \sum_{k=1}^{n} \mathbf{K}_k. \quad (A9)$$

**A2. The total static electric potential energy and the total magnetic potential energy in the body-fixed frame of $\beta_n$**

The total static electric potential energy and the total magnetic potential energy can be written relative to the body-fixed frame of $\beta_n$ as follows:

$$\begin{cases} U_{se} = -\sum_{k=1}^{n-1} \sum_{j=k+1}^{n} \dfrac{1}{4\pi\varepsilon_0} \int_{\beta_k} \int_{\beta_j} \dfrac{\rho_{se}(\mathbf{D}_k)\rho_{se}(\mathbf{D}_j) dV(\mathbf{D}_j) dV(\mathbf{D}_k)}{\|\mathbf{A}_n^T \mathbf{A}_k \mathbf{D}_k - \mathbf{A}_n^T \mathbf{A}_j \mathbf{D}_j + \mathbf{R}_{kn} - \mathbf{R}_{jn}\|} \\ U_m = -\sum_{k=1}^{n-1} \sum_{j=k+1}^{n} \dfrac{\mu_0}{4\pi} \int_{\beta_k} \int_{\beta_j} \dfrac{\mathbf{J}(\mathbf{D}_k) \cdot \mathbf{J}(\mathbf{D}_j) dV(\mathbf{D}_j) dV(\mathbf{D}_k)}{\|\mathbf{A}_n^T \mathbf{A}_k \mathbf{D}_k - \mathbf{A}_n^T \mathbf{A}_j \mathbf{D}_j + \mathbf{R}_{kn} - \mathbf{R}_{jn}\|} \end{cases}. \quad (A10)$$

If we denote $U = U_g + U_{se} + U_m$, the dynamical equation of the system is also Eq. (13).

The torque is

$$\begin{cases} \boldsymbol{\mu}^k = \boldsymbol{\mu}_g^k + \boldsymbol{\mu}_e^k + \boldsymbol{\mu}_m^k, & k = 1, 2, \cdots, n-1 \\ \boldsymbol{\mu}^n = \boldsymbol{\mu}_g^n + \boldsymbol{\mu}_{se}^n + \boldsymbol{\mu}_m^n \end{cases}, \quad (A11)$$

where

$$\begin{cases} \boldsymbol{\mu}_{se}^k = -\dfrac{1}{4\pi\varepsilon_0} \sum_{k=1}^{n-1} \sum_{j=k+1}^{n} \int_{\beta_k} \int_{\beta_j} \mathbf{A}_n^T \mathbf{A}_k \mathbf{D}_k \times \dfrac{\left(\mathbf{A}_n^T \mathbf{A}_k \mathbf{D}_k - \mathbf{A}_n^T \mathbf{A}_j \mathbf{D}_j + \mathbf{R}_{kn} - \mathbf{R}_{jn}\right)}{\left\| \mathbf{A}_n^T \mathbf{A}_k \mathbf{D}_k - \mathbf{A}_n^T \mathbf{A}_j \mathbf{D}_j + \mathbf{R}_{kn} - \mathbf{R}_{jn} \right\|^3} \rho_{se}(\mathbf{D}_k) \rho_{se}(\mathbf{D}_j) dV(\mathbf{D}_j) dV(\mathbf{D}_k) \\ \boldsymbol{\mu}_m^k = -\dfrac{\mu_0}{4\pi} \sum_{k=1}^{n-1} \sum_{j=k+1}^{n} \int_{\beta_k} \int_{\beta_j} \mathbf{A}_n^T \mathbf{A}_k \mathbf{D}_k \times \dfrac{\left(\mathbf{A}_n^T \mathbf{A}_k \mathbf{D}_k - \mathbf{A}_n^T \mathbf{A}_j \mathbf{D}_j + \mathbf{R}_{kn} - \mathbf{R}_{jn}\right)}{\left\| \mathbf{A}_n^T \mathbf{A}_k \mathbf{D}_k - \mathbf{A}_n^T \mathbf{A}_j \mathbf{D}_j + \mathbf{R}_{kn} - \mathbf{R}_{jn} \right\|^3} \mathbf{J}(\mathbf{D}_k) \cdot \mathbf{J}(\mathbf{D}_j) dV(\mathbf{D}_j) dV(\mathbf{D}_k) \end{cases}$$

(A12)

and

$$\begin{cases} \boldsymbol{\mu}_{se}^n = \dfrac{1}{4\pi\varepsilon_0} \sum_{k=1}^{n-1} \sum_{j=k+1}^{n} \int_{\beta_k} \int_{\beta_j} \mathbf{D}_n \times \dfrac{\left(\mathbf{A}_n^T \mathbf{A}_k \mathbf{D}_k - \mathbf{D}_n + \mathbf{R}_{kn}\right)}{\left\| \mathbf{A}_n^T \mathbf{A}_k \mathbf{D}_k - \mathbf{D}_n + \mathbf{R}_{kn} \right\|^3} \rho_{se}(\mathbf{D}_k) \rho_{se}(\mathbf{D}_j) dV(\mathbf{D}_j) dV(\mathbf{D}_k) \\ \boldsymbol{\mu}_m^n = \dfrac{\mu_0}{4\pi} \sum_{k=1}^{n-1} \sum_{j=k+1}^{n} \int_{\beta_k} \int_{\beta_j} \mathbf{D}_n \times \dfrac{\left(\mathbf{A}_n^T \mathbf{A}_k \mathbf{D}_k - \mathbf{D}_n + \mathbf{R}_{kn}\right)}{\left\| \mathbf{A}_n^T \mathbf{A}_k \mathbf{D}_k - \mathbf{D}_n + \mathbf{R}_{kn} \right\|^3} \mathbf{J}(\mathbf{D}_k) \cdot \mathbf{J}(\mathbf{D}_j) dV(\mathbf{D}_j) dV(\mathbf{D}_k) \end{cases}. \quad (A13)$$

Careful analysis of the motion of multiple-asteroid systems indicates that the static electric force and the magnetic force acting between each asteroid are negligible. Therefore, $U = U_g$ and $\begin{cases} \boldsymbol{\mu}^k = \boldsymbol{\mu}_g^k, & k = 1, 2, \cdots, n-1 \\ \boldsymbol{\mu}^n = \boldsymbol{\mu}_g^n \end{cases}$ are established for multiple-asteroid systems in the Solar system in Eqs. (13) and (15).

## Appendix B. Initial paramerers

Table 1. Initial conditions for calculation of dynamical configurations

*a) Major Body*

| Name of triple asteroid system | Mass (kg) | Bulk density (g·cm$^{-3}$) | Rotation period (h) | Pole solution in ecliptic EME2000 ($\lambda, \beta$) |
|---|---|---|---|---|
| (45) Eugenia[a1,a2,a3] | $5.62887 \times 10^{18}$ | 1.1 | 5.699 | (122.0, -19.2) |
| (87) Sylvia[b1,b2] | $1.478 \times 10^{19}$ | 1.29 | 5.18364 | (70.0, 69.0) |
| (93) Minerva[c] | $3.35 \times 10^{18}$ | 1.75 | 5.981767 | (11.0, 25.0) |
| (216) Kleopatra[d] | $4.64 \times 10^{18}$ | 3.6 | 5.385 | (76.0, 16.0) |
| (136617) 1994CC[e1,e2] | $2.66 \times 10^{11}$ | 2.1 | 2.3886 | (71.144, 67.486) |

*b)   Diameters and Moonlets*

| Name of triple asteroid system | Dimensions (km) | Moonlet 1 | | Moonlet 2 | |
|---|---|---|---|---|---|
| | | Name | Diameter (km) | Name | Diameter (km) |
| (45) Eugenia[a1,a2,a3] | 304×220×146 | Princesse | 5 | Petit_prince | 7 |
| (87) Sylvia[b1,b2] | 385×262×232 | Romulus | 10.8 | Remus | 10.6 |
| (93) Minerva[c] | 141.6 | Aegis | 3.6 | Gorgoneion | 3.2 |
| (216) Kleopatra[d] | 217×94×81 | Alexhelios | 8.9 | Cleoselene | 6.9 |
| (136617) 1994CC[e1,e2,e3] | 0.69×0.67×0.64 | Beta | 0.113 | Gamma | 0.08 |

[a1]Beauvalet et al. 2011. [a2]Beauvalet and Marchis 2014. [a3]Marchis et al. 2010. [b1]Berthier et al. 2014. [b2]Fang et al. 2012. [c]Marchis et al. 2013. [d]Descamps et al. 2010. [e1]Brozović et al. 2010. [e2]Brozović et al. 2011. [e3]Fang et al. 2011.

*c)   Initial Position and velocity for the moonlets of triple asteroid system (45) Eugenia [a3]*

| Bodies | R (km) | V (m/s) |
|---|---|---|
| Princesse | [198.698 -341.315 -519.467] | [-20.0509259 -11.580208 -0.2673796] |
| Petit-Prince | [772.798 548.518 664.893] | [11.7339120  0.1368715 -13.734375] |

*d)   Initial orbital parameters for the moonlets of triple asteroid system (87) Sylvia [b1]*

| *Orbital parameters* | Romulus | Remus |
|---|---|---|
| *Semi-major axis: a* (km) | 1357 | 706.5 |
| *Eccentricity: e* | 0.005566 | 0.02721 |
| *Inclination: i* (deg) | 8.293 | 7.824 |
| Long. of ascend. node: $\Omega$ (deg) | 92.6 | 94.8 |
| Arg. periapsis: $\omega$ (deg) | 61.06 | 357.0 |
| *Mean anomaly: M* (deg) | 197.0 | 261.0 |

*e)   Initial orbital parameters for the moonlets of triple asteroid system* (93) Minerva [c]

| *Orbital parameters* | Aegis | Gorgoneion |
|---|---|---|
| *Semi-major axis: a* (km) | 623.5 | 375.0 |
| *Eccentricity: e* | 0 | 0.05 |
| *Inclination: i* (deg) | 89.0 | 91.4 |
| Long. of ascend. node: $\Omega$ (deg) | 126.0 | 132.6 |
| Arg. periapsis: $\omega$ (deg) | 82.0 | 347.5 |
| *Mean anomaly: M* (deg) | 0.0 | 0.0 |

*f)   Initial orbital parameters for the moonlets of triple asteroid system* (216) Kleopatra [d]

| *Orbital parameters* | Alexhelios | Cleoselene |
|---|---|---|
| *Semi-major axis: a* (km) | 678.0 | 454.0 |
| *Eccentricity: e* | 0 | 0 |
| *Inclination: i* (deg) | 51.0 | 49.0 |
| Long. of ascend. node: $\Omega$ (deg) | 166.0 | 160.0 |
| Arg. periapsis: $\omega$ (deg) | 0 | 0 |

| | |
|---|---|
| *Mean anomaly: M* (deg) | 0 | 0 |

*g)  Initial orbital parameters for the moonlets of triple asteroid system* (136617) 1994CC[,e3]

| Orbital parameters | Beta | Gama |
|---|---|---|
| *Semi-major axis: a* (km) | 1.729 | 6.13 |
| *Eccentricity: e* | 0.002 | 0.192 |
| *Inclination: i* (deg) | 83.376 | 71.709 |
| Long. of ascend. node: $\Omega$ (deg) | 59.209 | 48.479 |
| Arg. periapsis: $\omega$ (deg) | 130.98 | 96.229 |
| *Mean anomaly: M* (deg) | 233.699 | 6.07 |

In Table 1c), 1e), 1f), and 1g), the orbital parameters are expressed respected to Earth's equator. In Table 1d), the orbital parameters are expressed respected to the primary's equator.

Table 2. Initial conditions for calculation of the dynamical configurations of the six-body system (134340) Pluto

*a)  Masses of the six-body system (134340) Pluto[f1,f2]*

| | Pluto | Charon | Nix | Hydra | Kerberos | Styx |
|---|---|---|---|---|---|---|
| Mass (kg) | 1.305e+22 | 1.52e+21 | 4.0e+17 | 8.0e+17 | 2.4e+16 | 5.25e+15 |
| Pole solution in ecliptic EME2000 ($\lambda, \beta$) (deg) | (132.993, -6.163) | (132.993, -6.163) | —— | —— | —— | —— |

*b)  Initial Position and velocity for the moonlets of six-body system (134340) Pluto[f1]*

| Bodies | R (km) | V (m/s) |
|---|---|---|
| Charon | [1297.17438478526 3752.60226174718 17011.90583845352] | [0.1453959508510873 0.1297771902069882 -0.0397230039699411] |
| Nix | [-30572.84277725838 -26535.81343448966 12311.29089587663] | [0.0232883188913688 0.0427977975396927 0.1464990283534413] |
| Hydra | [9024.34878023784 15210.73701650077 45591.75735722126] | [0.1004334400015913 0.0865524814427462 -0.0479498746416020] |
| Kerberos | [23564.20702505210 28380.03995076242 44578.02582182780] | [0.0792537025667567 0.0630220099842493 -0.0817084451068904] |
| Styx | [-43331.32611324427 -43628.45759453865 -20506.54193573317] | [-0.0374001037580065 -0.0184905610710285 0.1157937282701088] |

[f1]Brozović et al. 2015. [f2]Showalter and Hamilton. 2015